\documentclass[10pt,journal,compsoc]{IEEEtran}
\ifCLASSOPTIONcompsoc
  \usepackage[nocompress]{cite}
\else
  \usepackage{cite}
\fi
\ifCLASSINFOpdf
\else
\fi

\usepackage{multirow}
\usepackage{color}
\usepackage{bm}
\usepackage{indentfirst}
\usepackage[algoruled,linesnumbered]{algorithm2e}
\SetNlSty{textbf}{}{:}
\SetAlCapHSkip{0pt}% set caption margins
\SetAlgoHangIndent{0pt}
\makeatletter
\newcommand{\thealgorithm}{\arabic\algocf@float}
\newcommand{\AlgoCaptionFormat}{}

\renewcommand{\algocf@makecaption@ruled}[2]{%
  \global\sbox\algocf@capbox{\hskip\AlCapHSkip%
    \setlength{\hsize}{\columnwidth}% restored on exit of sbox
    \addtolength{\hsize}{-2\AlCapHSkip}% add equal margin to both sides
    \vtop{\AlgoCaptionFormat\algocf@captiontext{#1}{#2}}}% then caption is not centered
}%
\makeatother

\usepackage{xcolor} % http://www.ctan.org/tex-archive/macros/latex/contrib/xcolor
\usepackage{tikz}
\usetikzlibrary{fit,calc}
\newcommand*{\tikzmk}[1]{\tikz[remember picture,overlay,] \node (#1) {};\ignorespaces}
\newcommand{\boxit}[1]{\tikz[remember picture,overlay]{\node[yshift=2.8pt,xshift=3.7pt,fill=#1,opacity=.25,fit={(A)($(B)+(.878\linewidth,-2.4\baselineskip)$)}] {};}\ignorespaces}
\colorlet{pink}{red!40}
\colorlet{blue}{cyan!60}

\usepackage{xcolor}
\newcommand*\circled[1]{\tikz[baseline=(char.base)]{\node[font=\scriptsize,shape=circle,fill,inner sep=1pt] (char) {\textcolor{white}{#1}};}}

\usepackage{amsmath}
\usepackage{amssymb}

\DeclareFontFamily{U}{mathx}{}
\DeclareFontShape{U}{mathx}{m}{n}{<-> mathx10}{}
\DeclareSymbolFont{mathx}{U}{mathx}{m}{n}
\DeclareMathAccent{\widecheck}{0}{mathx}{"71}

\usepackage{comment}

\begin{document}
%
% paper title
% Titles are generally capitalized except for words such as a, an, and, as,
% at, but, by, for, in, nor, of, on, or, the, to and up, which are usually
% not capitalized unless they are the first or last word of the title.
% Linebreaks \\ can be used within to get better formatting as desired.
% Do not put math or special symbols in the title.
\title{Joint Linear and Nonlinear Computation across Functions
for Efficient Privacy-Preserving Neural Network Inference}
%
%
% author names and IEEE memberships
% note positions of commas and nonbreaking spaces ( ~ ) LaTeX will not break
% a structure at a ~ so this keeps an author's name from being broken across
% two lines.
% use \thanks{} to gain access to the first footnote area
% a separate \thanks must be used for each paragraph as LaTeX2e's \thanks
% was not built to handle multiple paragraphs
%
%
%\IEEEcompsocitemizethanks is a special \thanks that produces the bulleted
% lists the Computer Society journals use for "first footnote" author
% affiliations. Use \IEEEcompsocthanksitem which works much like \item
% for each affiliation group. When not in compsoc mode,
% \IEEEcompsocitemizethanks becomes like \thanks and
% \IEEEcompsocthanksitem becomes a line break with idention. This
% facilitates dual compilation, although admittedly the differences in the
% desired content of \author between the different types of papers makes a
% one-size-fits-all approach a daunting prospect. For instance, compsoc
% journal papers have the author affiliations above the "Manuscript
% received ..."  text while in non-compsoc journals this is reversed. Sigh.

\author{Qiao Zhang,Tao Xiang, Chunsheng Xin, Biwen Chen, and Hongyi Wu% <-this % stops a space
\IEEEcompsocitemizethanks{\IEEEcompsocthanksitem Qiao Zhang, Tao Xiang, and Biwen Chen are with College of Computer Science, Chongqing University, Chongqing, 400044, China. E-mail: \{qiaozhang, txiang, macrochen\}@cqu.edu.cn.
\IEEEcompsocthanksitem Chunsheng Xin is with Department of Electrical and Computer Engineering, Old Dominion University, Norfolk, VA, 23529, USA. E-mail: cxin@odu.edu.
\IEEEcompsocthanksitem Hongyi Wu is with Department of Electrical and Computer Engineering, University of Arizona, Tucson, AZ, 85721, USA. E-mail: mhwu@arizona.edu.}% <-this % stops an unwanted space
\thanks{Manuscript received April 19, 2005; revised August 26, 2015.}}

% note the % following the last \IEEEmembership and also \thanks -
% these prevent an unwanted space from occurring between the last author name
% and the end of the author line. i.e., if you had this:
%
% \author{....lastname \thanks{...} \thanks{...} }
%                     ^------------^------------^----Do not want these spaces!
%
% a space would be appended to the last name and could cause every name on that
% line to be shifted left slightly. This is one of those "LaTeX things". For
% instance, "\textbf{A} \textbf{B}" will typeset as "A B" not "AB". To get
% "AB" then you have to do: "\textbf{A}\textbf{B}"
% \thanks is no different in this regard, so shield the last } of each \thanks
% that ends a line with a % and do not let a space in before the next \thanks.
% Spaces after \IEEEmembership other than the last one are OK (and needed) as
% you are supposed to have spaces between the names. For what it is worth,
% this is a minor point as most people would not even notice if the said evil
% space somehow managed to creep in.

% The paper headers
\markboth{Journal of \LaTeX\ Class Files,~Vol.~14, No.~8, August~2015}%
{Shell \MakeLowercase{\textit{et al.}}: Bare Demo of IEEEtran.cls for Computer Society Journals}
% The only time the second header will appear is for the odd numbered pages
% after the title page when using the twoside option.
%
% *** Note that you probably will NOT want to include the author's ***
% *** name in the headers of peer review papers.                   ***
% You can use \ifCLASSOPTIONpeerreview for conditional compilation here if
% you desire.

% The publisher's ID mark at the bottom of the page is less important with
% Computer Society journal papers as those publications place the marks
% outside of the main text columns and, therefore, unlike regular IEEE
% journals, the available text space is not reduced by their presence.
% If you want to put a publisher's ID mark on the page you can do it like
% this:
%\IEEEpubid{0000--0000/00\$00.00~\copyright~2015 IEEE}
% or like this to get the Computer Society new two part style.
%\IEEEpubid{\makebox[\columnwidth]{\hfill 0000--0000/00/\$00.00~\copyright~2015 IEEE}%
%\hspace{\columnsep}\makebox[\columnwidth]{Published by the IEEE Computer Society\hfill}}
% Remember, if you use this you must call \IEEEpubidadjcol in the second
% column for its text to clear the IEEEpubid mark (Computer Society jorunal
% papers don't need this extra clearance.)

% use for special paper notices
%\IEEEspecialpapernotice{(Invited Paper)}

% for Computer Society papers, we must declare the abstract and index terms
% PRIOR to the title within the \IEEEtitleabstractindextext IEEEtran
% command as these need to go into the title area created by \maketitle.
% As a general rule, do not put math, special symbols or citations
% in the abstract or keywords.
\IEEEtitleabstractindextext{%
\begin{abstract}
While it is encouraging to witness the recent development in privacy-preserving Machine Learning as a Service (MLaaS), there still exists a significant performance gap for its deployment in real-world applications. We observe the state-of-the-art frameworks follow a compute-and-share principle for every function output where the summing in linear functions, which is the last of two steps for function output, involves all rotations (which is the most expensive HE operation), and the multiplexing in nonlinear functions, which is also the last of two steps for function output, introduces noticeable communication rounds. Therefore, we challenge the conventional compute-and-share logic and introduce the first joint linear and nonlinear computation across functions that features by 1) \textit{the PHE triplet} for computing the nonlinear function, with which the multiplexing is eliminated; 2) \textit{the matrix encoding} to calculate the linear function, with which all rotations for summing is removed; and 3) \textit{the network adaptation} to reassemble the model structure, with which the joint computation module is utilized as much as possible. The boosted efficiency is verified by the numerical complexity, and the experiments demonstrate up to 13$\times$ speedup for various functions used in the state-of-the-art models and up to 5$\times$ speedup over mainstream neural networks.
\end{abstract}

% Note that keywords are not normally used for peerreview papers.
}

% make the title area
\maketitle

% To allow for easy dual compilation without having to reenter the
% abstract/keywords data, the \IEEEtitleabstractindextext text will
% not be used in maketitle, but will appear (i.e., to be "transported")
% here as \IEEEdisplaynontitleabstractindextext when the compsoc
% or transmag modes are not selected <OR> if conference mode is selected
% - because all conference papers position the abstract like regular
% papers do.
\IEEEdisplaynontitleabstractindextext
% \IEEEdisplaynontitleabstractindextext has no effect when using
% compsoc or transmag under a non-conference mode.

% For peer review papers, you can put extra information on the cover
% page as needed:
% \ifCLASSOPTIONpeerreview
% \begin{center} \bfseries EDICS Category: 3-BBND \end{center}
% \fi
%
% For peerreview papers, this IEEEtran command inserts a page break and
% creates the second title. It will be ignored for other modes.
\IEEEpeerreviewmaketitle

\IEEEraisesectionheading{
\section{Introduction}\label{sec:introduction}}
\IEEEPARstart{D}{eep} Learning (DL) has dramatically evolved in the recent decade \cite{lecun1998gradient,krizhevsky2012imagenet,simonyan2014very,he2016deep} and been successfully deployed in many applications such as image classification~\cite{szegedy2015going}, voice recognition~\cite{zhang2020pushing} and financial evaluation~\cite{sohangir2018big}. Due to the need for massive training data and
adequate computation resources~\cite{najafabadi2015deep}, it is often impractical for an end user (the client $\mathcal{C}$) to train her DL model. To this end, the Machine Learning as a Service (MLaaS) emerges as a feasible alternative where a server $\mathcal{S}$ in the cloud owns a neural network that is well trained on plenty of data, and the client uploads her input to the server to obtain the prediction result.

However, noticeable privacy concerns raise in MLaaS since the client must send her private data, which could be sensitive, to the server. In many cases, the client prefers to obtain the prediction without letting other parties, including the cloud server, know her data. In fact, regulations have been enforced to forbid the disclosure of private data, e.g., the Health Insurance Portability and Accountability Act (HIPAA)~\cite{assistance2003summary} for medical data and the General Data Protection Regulation (GDPR)~\cite{goddard2017eu} for business data. Meanwhile, the cloud server intends to hide the proprietary parameters of its well-trained neural network, and only return the model output in response to the client's prediction request.

\textcolor{black}{Privacy-preserving MLaaS takes both legal and ethical concerns for the client's private data and the server's model parameters into account, which aims to ensure that 1) the server learns nothing about the client's private data and 2) the client learns nothing about the server's model parameters beyond what can be returned from the network output, e.g., the predicted class. The key challenge in privacy-preserving MLaaS is how to efficiently embed cryptographic primitives into function computation of neural networks, which otherwise may lead to prohibitively high computation complexity and/or degraded prediction accuracy due to large-size circuits and/or function approximations.}

To achieve usable privacy-preserving MLaaS, a series of recent works have made inspiring progress towards the system efficiency~\cite{gilad2016cryptonets,liu2017oblivious,juvekar2018gazelle,mohassel2017secureml,riazi2019xonn,mishra2020delphi,rathee2020cryptflow2,zhang2021gala,patra2021aby2,tan2021cryptgpu,riazi2018chameleon,rouhani2018deepsecure,mohassel2018aby3,boemer2020mp2ml,demmler2015aby}. Specifically, the inference speed has gained several orders of magnitude from CryptoNets~\cite{gilad2016cryptonets} to the recent frameworks. At a high level, these privacy-preserving frameworks carefully consider and adopt several cryptographic primitives (e.g., the Homomorphic Encryption (HE)~\cite{brakerski2012fully,fan2012somewhat,cheon2017homomorphic} and Multi-Party Computation (MPC) techniques~\cite{goldreich1998secure} (such as Oblivious Transfer (OT)~\cite{brassard1986all}, Secret Sharing (SS)~\cite{shamir1979share} and Garbled Circuits (GC)~\cite{bellare2012foundations,yao1986generate})) to compute the linear (e.g., dot product and convolution) and nonlinear (e.g., \textsf{ReLU}) functions, which are repeatedly stacked and act as the building blocks in a neural network.
Among them, the mixed-protocol approaches that utilizes HE to compute linear functions while adapting MPC for nonlinear functions show more efficiency advantages for the privacy-preserving MLaaS with two-party computation~\cite{liu2017oblivious,juvekar2018gazelle,rathee2020cryptflow2}.
For example, CrypTFlow2~\cite{rathee2020cryptflow2} has shown significant speedup compared with other state-of-the-art schemes such as GAZELLE~\cite{juvekar2018gazelle} %(USENIX Security'18)
and DELPHI~\cite{mishra2020delphi}.

While it is encouraging to witness the recent development in privacy-preserving MLaaS, there still exists a significant performance gap for its deployment in real-world applications. For example,
our benchmark has shown that CrypTFlow2 takes 115 seconds and 147 seconds to run the well-known DL networks \texttt{VGG-19}~\cite{simonyan2014very} and \texttt{ResNet-34}~\cite{he2016deep} on the Intel(R) Xeon(R) E5-2666 v3 $@$ 2.90GHz CPU (see the detailed experimental settings and results in Sec.~\ref{evaluation}). It is worth pointing out that the constraints of response time in many practical ML-driven applications (such as speech recognition and wearable health monitoring) are within a few seconds or up to one minute~\cite{Alexa,mohammadzadeh2018prediction}. This performance gap motivates us to further improve the efficiency of privacy-preserving MLaaS especially for the mixed-protocol approaches.

We begin with analyzing the computation logic in these state-of-the-art privacy-preserving frameworks. Concretely, the basic logic in all designs is to firstly calculate the output of a function, based on specific cryptographic primitives. That securely computed output is in an ``encrypted'' form and is then shared between $\mathcal{C}$ and $\mathcal{S}$. The respective share at $\mathcal{C}$ and $\mathcal{S}$ acts as the input of next function. As a neural network consists of stacked linear and nonlinear functions, this compute-and-share logic for each function output is sequentially repeated until the last one. For example, the initial input of CrypTFlow2 is $\mathcal{C}$'s private data, which is encrypted and sent to $\mathcal{S}$. $\mathcal{S}$ conducts HE-based computation for the linear function where the HE addition, multiplication, and rotation (which are three basic operators over encrypted data) are performed between $\mathcal{C}$-encrypted data and $\mathcal{S}$'s model parameters.
It produces an encrypted function output which is then shared (in plaintext) between $\mathcal{C}$ and $\mathcal{S}$, and those shares serve as the input of the following OT-based computation for subsequent nonlinear function, whose corresponding output shares act as the input for the next function. This computation mode is repeated in function wise until $\mathcal{C}$ gets the network output.

While this compute-and-share mode for each function output is seemly logical and all of the aforementioned works follow this principle, we observe the fact that the summing in linear functions, which is the last of two steps (where the first step is multiplication) to get the function output, involves all rotations (which is the most expensive HE operation), and the multiplexing in nonlinear functions, which is also the last of two steps (where the first step is
comparison) to get the function output, introduces noticeable communication rounds (e.g., about one third in OT-based communication rounds\footnote{One round is the communication trip from source node to sink node and then from sink node back to source node, and 0.5 round is either communication trip from source node to sink node or the one from sink node to source node.}~\cite{rathee2020cryptflow2}). Therefore we ask the following natural question:

\vspace*{0.05in}
\textit{\underline{Is it possible to efficiently circumvent the compute-and-share} \underline{logic for the function output such that we can evade expensive r-} \underline{otations and noticeable communication rounds at the last step of} \underline{function computation to achieve more efficient privacy-preservi-\,} \underline{ng MLaaS?}}

\vspace*{0.05in}
\textbf{Our Contributions.} In this paper, we give an affirmative answer to this question. To this end, we challenge and break the conventional compute-and-share logic for function output, and propose the new \textit{share-in-the-middle logic} towards function computation for efficient privacy-preserving MLaaS. In particular, we come up with the first joint linear and nonlinear computation across functions where the expensive rotations and noticeable communication rounds at the last step of function computation are efficiently removed, via a careful utilization of the intermediates during the function computation. The underlying bases of our joint computation across functions feature with the following three novel designs:
\begin{enumerate}
  %\vspace*{-0.08in}
  \item\textit{the PHE triplet} for computing the intermediates of nonlinear function, with which the communication cost for multiplexing is totally eliminated. Meanwhile, it is not only offline-oriented (i.e., its generation is completely independent of $\mathcal{C}$'s private input) but also non-interactive (i.e., its generation is fully asynchronous between $\mathcal{C}$ and $\mathcal{S}$), which is much more flexible compared with other offline-but-synchronous generation~\cite{liu2017oblivious,mishra2020delphi}.
  %\vspace*{-0.1in}
  \item\textit{the matrix encoding} to calculate the intermediates of linear function, with which all rotations for summing is totally removed. For instance, 2048 rotations are needed to calculate one of the convolutions in ~\texttt{ResNet}~\cite{he2016deep} while we save this cost via the proposed matrix encoding. Furthermore, the PHE triplet is integrated with the matrix encoding to form our joint computation block (to be discussed in Sec.~\ref{sys:jointcom}), which shows better efficiency from both numerical and experimental analysis.
  \item\textit{the network adaptation} to reassemble the DL architecture (e.g., adjust the function orders and decompose the function computation to make it integrated with both previous and subsequent functions) such that the recombined functions are more suitable for applying the proposed PHE triplet and matrix encoding, and thus can take more advantages of our proposed joint computation block, which finally contributes to the overall efficiency improvement.
\end{enumerate}

The boosted efficiency of our protocol is demonstrated by both the numerical complexity analysis and the experimental results. For example, we achieve up to 13$\times$ speedup for various functions used in the state-of-the-art DL models and up to $5\times$ speedup over mainstream neural networks, compared with the state-of-the-art privacy-preserving frameworks. Furthermore, we lay our work as an initial attempt for the share-in-the-middle computation, which could inspire more works in this new lane for efficient privacy-preserving MLaaS.

On the other hand, we also notice two most recent works that demonstrate comparable efficiency as ours. The first one is COINN \cite{hussain2021coinn} which features with the well-designed model customization and ciphertext execution, and our computation can be built on top of COINN's quantized models to gain more speedup. The second one is Cheetah~\cite{huang2022cheetah} which utilizes the accumulative property of polynomial multiplication and the ciphertex extraction to eliminate rotations and relies on VOLE-style OT to boost the nonlinear computation. While we can also benefit from the optimized OT to compute nonlinear functions, we differentiate our work from Cheetah's rotation elimination by exploiting the accumulative nature in matrix-vector multiplication where the output can be viewed as the linear combination of all columns in the weight matrix.
Furthermore, both COINN and Cheetah are under the compute-and-share logic while we set our work as the first one for share-in-the-middle computation.

The rest of the paper is organized as follows. In Sec.~\ref{preliminary}, we introduce the system setup and the primitives that are adopted in our protocol. Sec.~\ref{sysdescrib} elaborates the design of our joint linear and nonlinear computation. The experimental results are illustrated and discussed in Sec.~\ref{evaluation}. %Section~\ref{related_work} gives the related work of privacy-preserving DL.
Finally, we conclude the paper in Sec.~\ref{conclusion}.

\section{Preliminaries}\label{preliminary}
\indent\textbf{Notations.} We denote $[k]$ as the set of integers $\{0,1,\dots,k-1\}$.
$\lceil\;\rceil$ and $\left \lfloor\;\right \rfloor$ denote the ceiling and flooring function, respectively.
Let \textbf{1}$\{\mathcal{I}\}$ denote the indicator function that is 1 when $\mathcal{I}$ is true and 0 when $\mathcal{I}$ is false. $r$ $\scriptstyle\overset{\$}{\gets}$ $\mathcal{D}$ denotes randomly sampling a component $r$ from a set $\mathcal{D}$. $``\cdots,(i_1:i_2),\cdots"$ includes indices from $i_1$ to $(i_2-1)$ for one dimension while $``\cdots,:,\cdots"$ contains all indices in that dimension. $``|"$ represents the concatenation of matrices or numbers.

\begin{figure}[!tbp]
\centering
\includegraphics[trim={5.8cm 3.9cm 8.3cm 11.2cm}, clip, scale=0.59]{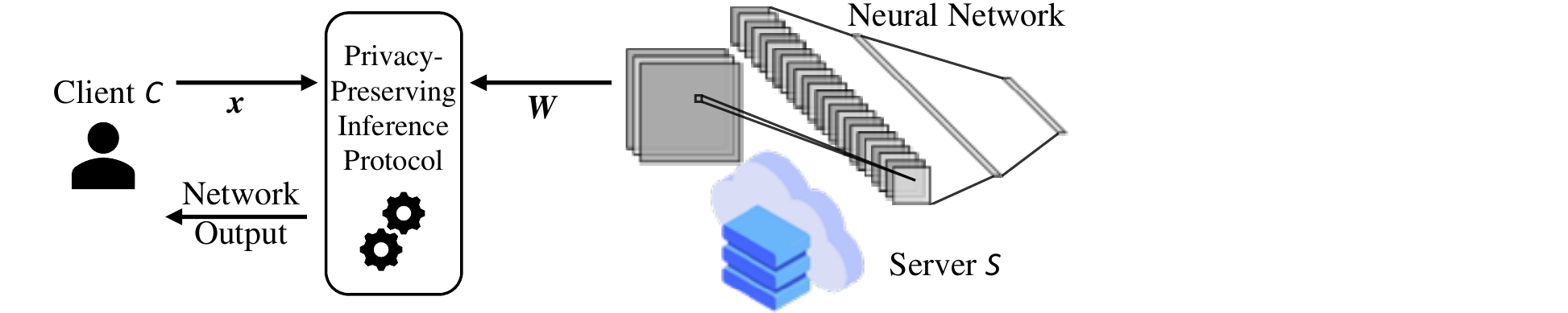}
\vspace*{-0.25in}
\caption{Overview of Cryptographic Inference.}%\vspace*{-0.1in}
\label{fig:mlaas}
\vspace*{-0.15in}
\end{figure}
\subsection{System Model}
\textcolor{black}{We consider the context of cryptographic inference (as shown in Figure~\ref{fig:mlaas}) where $\mathcal{C}$ holds a private input $\bm{x}$ and $\mathcal{S}$ holds the neural network with proprietary model parameters $\bm{W}$. After the inference, $\mathcal{C}$ learns two pieces of information: the network architecture (such as the number, types and dimensions of involved functions) and the network output\footnote{Note that this learnt information is commonly assumed in the state-of-the-art frameworks such as MiniONN~\cite{liu2017oblivious} and Cheetah~\cite{huang2022cheetah}.}, while $\mathcal{S}$ learns nothing. The neural network processes $\bm{x}$ through a sequence of linear and nonlinear functions to finally classify $\bm{x}$ into one of the potential
classes. Specifically, we target at the widely-applied Convolutional Neural Network (CNN) and describe its included functions as follows.}

\textbf{Convolution (\textsf{Conv}).}
The \textsf{Conv} operates between a three-dimension input $\bm{a}\in\mathcal{R}^{c_i\times{h_i}\times{w_i}}$ (the $\mathcal{R}$ is defined in Sec. 2.3.1) and the kernel $\textbf{K}\in\mathcal{R}^{c_o\times{c_i}\times{f_h}\times{f_w}}$ with a stride $s\in\mathbb{N}^+$ where $c_i$ and $c_o$ are the number of input and output channels, $h_i$ and $w_i$ are the height and width of each two-dimension input channel, and $f_h$ and $f_w$ are the height and width (which are always equal) of each two-dimension filter of \textbf{K}. \textcolor{
black}{The output is a three-dimension matrix} $\bm{y}=\textsf{Conv}(\bm{a},\textbf{K})\in\mathcal{R}^{c_o\times{h_i'}\times{w_i'}}$ where $h_i'=\left \lceil \frac{h_i}{s} \right \rceil$ and $w_i'=\left \lceil \frac{w_i}{s} \right \rceil$ are the height and width of the two-dimension output channel. Such strided convolution is mathematically expressed as
\[
\bm{y}_{\alpha,\beta,\gamma}=\sum_{\lambda\in{[c_i]}}\sum_{\ell_1,\ell_2\in\bm{\delta}}\bm{a}_{\lambda,s\beta+\ell_1,s\gamma+\ell_2}\textbf{K}_{\alpha,\lambda,\ell_1+\Delta,\ell_2+\Delta}
\]
where $\bm{\delta}=\mathbb{Z}\cap[-\frac{f_h-1}{2},\frac{f_h-1}{2}]$, $\Delta=\frac{f_h-1}{2}$ and $\bm{a}_{\lambda,s\beta+\ell_1,s\gamma+\ell_2}=0$ if $s\beta+\ell_1<0$ or $s\gamma+\ell_2<0$. The summing process in the cryptographic inference inevitably introduces a series of expensive rotations which limits the inference efficiency, a carefully designed matrix encoding is proposed in this work, together with the joint computation with the nonlinear function,
%namely \textsf{ReLU},
which efficiently removes all the expensive rotations (see Sec.~\ref{sysdescrib}).

\textbf{Dot Product (\textsf{Dot}).}
The input to the dot product is a $n_i$-sized vector $\bm{a}\in\mathcal{R}^{n_i}$ and the output is the $n_o$-sized vector $\bm{y}=\textsf{Dot}(\textbf{W},\bm{a})\in\mathcal{R}^{n_o}$ where
\begin{equation}\label{yj}
\bm{y}_{j}=\sum_{\lambda\in{[n_i]}}\textbf{W}_{j,\lambda}\bm{a}_{\lambda}
\end{equation}
and $\textbf{W}\in\mathcal{R}^{n_o\times{n_i}}$. As the dot product and convolution are both intrinsically weighted sums, we treat dot product similarly with convolution in the cryptographic inference where the proposed matrix encoding for \textsf{Dot}, along with the joint computation with the nonlinear function (see Sec.~\ref{sysdescrib}), efficiently eliminates all the expensive rotations.

\textbf{Batch Normalization (\textsf{BN}).}
In the neural network inference, the \textsf{BN} scales and shifts each two-dimension input channel by a constant $\bm{\mu}_{\beta}\in\mathcal{R}$ and $\bm{\theta}_{\beta}\in\mathcal{R}$, respectively, given the input $\bm{a}\in\mathcal{R}^{c_i\times{h_i}\times{w_i}}$. This operation is mathematically expressed as
\[
\bm{a}_{\beta,:,:}\gets\bm{\mu}_{\beta}\bm{a}_{\beta,:,:}+\bm{\theta}_{\beta}
\]

As the \textsf{BN} always follows behind the convolution, we integrate it with the convolution (see Sec.~\ref{sysdescrib}), together with our joint computation with nonlinear function, to improve the efficiency for calculating the $\textsf{Conv+BN}$.

\textbf{\textsf{ReLU}.}
For a value $a\in\mathcal{R}$, the \textsf{ReLU} is calculated as $\textsf{ReLU}(a)=a\cdot\textbf{1}\{a\}$. In the context of cryptographic inference, the multiplication between $a$ and $\textbf{1}\{a\}$ (denoted by $\textsf{DReLU}(a)$ thereafter) involves the multiplexing with noticeable communication rounds (e.g., about one third for OT-based communication rounds in~\cite{rathee2020cryptflow2}). In this work, we combine the \textsf{ReLU} computation with \textsf{Conv} (see Sec.~\ref{sysdescrib}) to totally relieve that cost.

\textbf{Mean Pooling (\textsf{MeanPool}).} Given the input $\bm{a}\in\mathcal{R}^{c_i\times{h_i}\times{w_i}}$, the \textsf{MeanPool} sums and averages the components in each $s_{\textsf{mp}}\times{s_{\textsf{mp}}}$ pooling window where $s_{\textsf{mp}}\in\mathbb{N}^+$, and returns all mean values as output. In this way, the output size becomes $c_i\times{\left \lceil \frac{h_i}{s_{\textsf{mp}}} \right \rceil}\times{\left \lceil \frac{w_i}{s_{\textsf{mp}}} \right \rceil}$. Since the \textsf{MeanPool} always appears between $\textsf{Conv}$ and $\textsf{ReLU}$, we decouple its summing and averaging to $\textsf{ReLU}$ and $\textsf{Conv}$ (see Sec.~\ref{sysdescrib}) to utilize our joint computation block for a more efficient cryptographic inference.

\textbf{Max Pooling  (\textsf{MaxPool}).}
The \textsf{MaxPool} works similarly as \textsf{MeanPool} except that the returned value is the maximum in each $s_{\textsf{mxp}}\times{s_{\textsf{mxp}}}$ pooling window where $s_{\textsf{mxp}}\in\mathbb{N}^+$. The state-of-the-art framework deals with \textsf{MaxPool} in each $s_{\textsf{mxp}}\times{s_{\textsf{mxp}}}$ pooling window via $(s_{\textsf{mxp}}^2-1)$ sequential comparisons~\cite{rathee2020cryptflow2} and we further parallel this process by considering the independent nature of two comparisons among four values (see Sec.~\ref{sysdescrib}).

\textbf{\textsf{ArgMax}.}
The \textsf{ArgMax} is usually the last function in a neural network to classify $\mathcal{C}$'s input to a potential class. \textsf{ArgMax} operates similarly as the \textsf{MaxPool} except that its pooling window embraces all values of the input and it finally returns the index of the maximum. The calculation of \textsf{ArgMax} is traditionally in a sequential manner~\cite{rathee2020cryptflow2}, which is optimized in a similar way as \textsf{MaxPool}.

\subsection{Threat Model}
Our protocol involves two parties namely the client $\mathcal{C}$ and the server $\mathcal{S}$, and we follow in this work the static semi-honest security definition~\cite{lindell2017simulate} for secure two-party computation.

\textbf{Static Semi-Honest Security.} There are two parties denoted by $\mathcal{C}$ and $\mathcal{S}$. Let $f_{\mathcal{C}}(\bm{x}, \bm{W})$ and $f_{\mathcal{S}}(\bm{x}, \bm{W}))$ be the output for $\mathcal{C}$ and $\mathcal{S}$ in the ideal functionality $\mathcal{F}$, respectively, while $f(\bm{x}, \bm{W})=(f_{\mathcal{C}}(\bm{x}, \bm{W}),f_{\mathcal{S}}(\bm{x}, \bm{W})))$ be the joint output. Let the \textsf{view} of $\mathcal{C}$ and $\mathcal{S}$ during an execution of $\Pi$ on inputs $(\bm{x}, \bm{W})$ be $\textsf{view}_{\mathcal{C}}^{\Pi}(\bm{x}, \bm{W})$ and $\textsf{view}_{\mathcal{S}}^{\Pi}(\bm{x}, \bm{W})$ that consist of the private input $\bm{x}$ and model parameters $\bm{W}$, as well as the contents of internal random tap and the messages received at $\mathcal{C}$ and $\mathcal{S}$ during the execution, respectively. Similarly, $\textsf{output}_{\mathcal{C}}^{\Pi}(\bm{x}, \bm{W})$ and $\textsf{output}_{\mathcal{S}}^{\Pi}(\bm{x}, \bm{W})$ are the output of $\mathcal{C}$ and $\mathcal{S}$ during an execution of $\Pi$ on inputs $(\bm{x}, \bm{W})$ and can be computed from the $\Pi$'s view. The joint output of both parties is $\textsf{output}^{\Pi}(\bm{x}, \bm{W})=(\textsf{output}_{\mathcal{C}}^{\Pi}(\bm{x}, \bm{W}),\textsf{output}_{\mathcal{S}}^{\Pi}(\bm{x}, \bm{W}))$.

\vspace*{0.1in}
\noindent\textbf{Definition 1.} \textit{A protocol $\Pi$ securely computes $\mathcal{F}$ against static semi-honest adversaries $\mathcal{A}$ if there exist probabilistic polynomial-time (PPT) algorithms} $\textsf{Sim}_{\mathcal{C}}$ \textit{and} $\textsf{Sim}_{\mathcal{S}}$ \textit{such that}
{\footnotesize\[
\{\textsf{Sim}_{\mathcal{C}}(\bm{x},f_{\mathcal{C}}(\bm{x}, \bm{W})),f(\bm{x}, \bm{W})\}\overset{\textsf{c}}{\equiv}\{\textsf{view}_{\mathcal{C}}^{\Pi}(\bm{x}, \bm{W}),\textsf{output}^{\Pi}(\bm{x}, \bm{W})\},
\]
\[
\{\textsf{Sim}_{\mathcal{S}}(\bm{W},f_{\mathcal{S}}(\bm{x}, \bm{W})),f(\bm{x}, \bm{W})\}\overset{\textsf{c}}{\equiv}\{\textsf{view}_{\mathcal{S}}^{\Pi}(\bm{x}, \bm{W}),\textsf{output}^{\Pi}(\bm{x}, \bm{W})\}.
\]}
\quad The main ideal functionalities for our protocol are  $\mathcal{F}_{\textsf{DReLU}}$ and $\mathcal{F}_{\textsf{ReConv}}$ as shown in  Algorithm~\ref{alg:ideal_drelu} and Algorithm~\ref{alg:ideal_reconv}, and we prove in Sec. 3.2.5 that our designed protocol $\Pi_{\textsf{ReConv}}^{\textrm{ring},p}$ presented in Algorithm~\ref{alg:reconv} securely realizes $\mathcal{F}_{\textsf{ReConv}}$ in the $\{\mathcal{F}_{\textsf{DReLU}}\}$-hybrid model, in the presence of semi-honest adversaries. Furthermore, we do not defend against attacks, such as the API attacks~\cite{shokri2017membership,tramer2016stealing}, which are based purely on the inference results and other orthogonal techniques, such as differential privacy~\cite{abadi2016deep,jayaraman2019evaluating}, can be utilized to provide more privacy guarantee~\cite{huang2022cheetah}.

\subsection{Cryptographic Primitives}\label{cryptoprimit}
\noindent\textbf{2.3.1 Packed Homomorphic Encryption (PHE).}
PHE is a primitive that supports
various vector operations over encrypted data without decryption, and generates an encrypted result which matches the corresponding operations on plaintext~\cite{brakerski2012fully,fan2012somewhat,cheon2017homomorphic}. Specifically, given a vector $\bm{x} =(x_0,x_1,\dots,x_{n-1})\in \mathcal{R}^n$ over a ring $\mathcal{R}=\mathbb{Z}_{p}=\mathbb{Z}\cap(-p/2,p/2]$, it is encrypted into a ciphertext \textsf{ct}$_{\bm{x}}^{\textsf{pk}_{\mathcal{C}}}=$ \textsf{Enc}$^{\textsf{pk}_{\mathcal{C}}}(\bm{x})$ where \textsf{pk}$_{\mathcal{C}}$ denotes the public key of the client $\mathcal{C}$ and we thereafter use \textsf{ct}$_{\bm{x}}^{{\mathcal{C}}}=$ \textsf{Enc}$^{{\mathcal{C}}}(\bm{x})$ for brevity. Similar logic is applied for the ciphertext \textsf{ct}$_{\bm{x}}^{{\mathcal{S}}}=$ \textsf{Enc}$^{{\mathcal{S}}}(\bm{x})$ where $\bm{x}$ is encrypted by the public key of the server $\mathcal{S}$. The correctness of PHE is firstly guaranteed by a decryption process such that $\bm{x} =$ \textsf{Dec}$^{\textsf{sk}_{\star}}(\textsf{ct}_{\bm{x}}^{\star})$ where ${\star}\in\{\mathcal{C},\mathcal{S}\}$, $\textsf{sk}_{\star}$
is the secret key of either $\mathcal{C}$ or $\mathcal{S}$, and \textsf{Dec}$^{\star}()$ is used henceforth to simplify the notation for decryption. Second, the PHE system can securely evaluate an arithmetic circuit consisting of addition and multiplication
gates by leveraging the following operations where $\bm{u}=(u_0,\dots,u_{n-1})$ (resp. $\bm{v}=(v_0,\dots,v_{n-1})$) $\in\mathcal{R}^n$, and $``+"$ (resp. $``-"$ and $``\cdot"$) is the component-wise addition (resp. subtraction and multiplication).
\begin{itemize}
  \item Homomorphic addition $(\oplus)$ and subtraction $(\ominus)$: \textsf{Dec}$^{\star}(\textsf{ct}_{\bm{u}}^{\star}$ $\oplus$ $\textsf{ct}_{\bm{v}}^{\star})=\bm{u}+\bm{v}$ (resp. \textsf{Dec}$^{\star}(\textsf{ct}_{\bm{u}}^{\star}\oplus{\bm{v}})=\bm{u}+\bm{v}$) and \textsf{Dec}$^{\star}(\textsf{ct}_{\bm{u}}^{\star}\ominus\textsf{ct}_{\bm{v}}^{\star})=\bm{u}-\bm{v}$ (resp. \textsf{Dec}$^{\star}(\textsf{ct}_{\bm{u}}^{\star}\ominus{\bm{v}})=\bm{u}-\bm{v}$).
  \item Homomorphic multiplication $(\otimes)$: \textsf{Dec}$^{\star}(\textsf{ct}_{\bm{u}}^{\star}\otimes\textsf{ct}_{\bm{v}}^{\star})=\bm{u}\cdot\bm{v}$ (resp. \textsf{Dec}$^{\star}(\textsf{ct}_{\bm{u}}^{\star}\otimes{\bm{v}})=\bm{u}\cdot\bm{v}$).
  \item Homomorphic rotation (\textsf{Rot}): \textsf{Dec}$^{\star}(\textsf{Rot}(\textsf{ct}_{\bm{u}}^{\star};\ell))=\bm{u}_{\ell}$ where $\bm{u}_{\ell}=\rho(\bm{u};\ell)=(u_{\ell},\dots,u_{n-1},u_0,\dots,u_{\ell-1}$ $)$ and $\ell\in[n]$. Note that a rotation by $(-\ell)$ is the same as a rotation by $(n-\ell)$.
\end{itemize}

Given the above four operations, the runtime complexity of \textsf{Rot} is significantly larger than that of $\oplus$, $\ominus$ and $\otimes$~\cite{huang2022cheetah,boemer2020mp2ml}. Meanwhile, the runtime complexity of homomorphic multiplication between two ciphertext namely $\textsf{ct}_{\bm{u}}^{\star}$ $\otimes$ $\textsf{ct}_{\bm{v}}^{\star}$ is also larger than that of homomorphic multiplication between one ciphertext and one plaintext namely $\textsf{ct}_{\bm{u}}^{\star}$ $\otimes$ ${\bm{v}}$. While the neural network is composed of a stack of linear and nonlinear functions, the PHE is preferably utilized to calculate the linear functions such as \textsf{Dot} and \textsf{Conv} in the privacy-preserving neural networks, due to its intrinsic support for linear arithmetic computation~\cite{demmler2015aby}. Such computation process involves calling aforementioned PHE operations to get the encrypted output of linear function, which is then shared between $\mathcal{C}$ and $\mathcal{S}$ to serve as the input for computing next function, and this compute-and-share logic for the output of linear function is undoubtedly adopted in the state-of-the-art privacy-preserving frameworks~\cite{juvekar2018gazelle,mishra2020delphi}.

\textcolor{black}{Unfortunately, this compute-and-share process requires a series of expensive \textsf{Rot} due to the needed summing in linear functions~\cite{juvekar2018gazelle}, and the PHE based linear calculation dominates the overall cost of neural model computation~\cite{zhang2021gala}. Therefore it makes the rotation elimination a significant and challenging task to achieve the next-stage boost towards computation efficiency. In this work, we jointly consider the computation for both linear and nonlinear functions, which enables us to feed the specifically-
formed PHE intermediates from the linear function to the computation for the subsequent nonlinear function. Such joint design, rather than traditionally separate computation for each function output, contributes to efficiently eliminate all \textsf{Rot}\footnote{Our design also involves no ciphertext-ciphertext but faster ciphertext-plaintext multiplication.} in the entire process of cryptographic inference.}

\noindent\textbf{2.3.2 Oblivious Transfer (OT).}
\textcolor{black}{In this work}, we rely on OT to compute the comparison-based nonlinear functions such as \textsf{ReLU}, \textsf{MaxPool} and \textsf{ArgMax}. Specifically, in the 1-out-of-$k$ OT, $( \mathop{}_{1}^{k})$-OT$_{\ell}$, the sender inputs $k$ $\ell$-length strings $m_0,\dots,m_{k-1}$, and the receiver's input is a value $i\in[k]$. The receiver gets $m_i$ after the OT functionality while the sender obtains nothing~\cite{kolesnikov2013improved}. The state-of-the-art computation for a comparison-based nonlinear function mainly involves the OT-based \textsf{DReLU} functionality along with the OT-based multiplexer (\textsf{Mux}) to get the corresponding shares of such function output, which act as the input to compute next function~\cite{rathee2020cryptflow2}.
While such process has relatively light computation and transmission load, the \textsf{Mux} accounts for nearly one third of the total communication rounds~\cite{rathee2020cryptflow2}, which hinders the further optimization towards overall efficiency.
In this paper, we totally remove the required \textsf{Mux} by a joint computation for both nonlinear and subsequent (linear) functions. Instead of getting the output shares with \textsf{Mux},
our design directly utilizes the nonlinear intermediates (e.g., \textsf{DReLU}) to generate the input of subsequent function.

\noindent\textbf{2.3.3 Secret Sharing (SS).}
The additive SS over the ring $\mathcal{R}=\mathbb{Z}_{p}$ is adopted in this paper to form the output shares of linear or nonlinear functions, which serve as the input to compute next function. Here $p$ is either the plaintext modulus determined by the PHE for the linear functions or two for the boolean computation of nonlinear functions. Specifically, the SS algorithm $\textsf{Shr}(x)=(\langle x\rangle^{\mathcal{C}}_p,\langle x\rangle^{\mathcal{S}}_p)$ inputs an element $x$ in $\mathbb{Z}_{p}$ and generates shares of $x$, $\langle x\rangle^{\mathcal{C}}_p$ and $\langle x\rangle^{\mathcal{S}}_p$, over $\mathbb{Z}_{p}$. The shares are randomly sampled in $\mathbb{Z}_{p}$ such that $\langle x\rangle^{\mathcal{C}}_p\boxplus\langle x\rangle^{\mathcal{S}}_p=x$ where $``\boxplus"$ is the addition over $\mathbb{Z}_{p}$ and we similarly denote $``\boxminus"$ and $``\boxdot"$ as the subtraction and multiplication over $\mathbb{Z}_{p}$.
Furthermore, we denote the reconstruction of $x$ as $\textsf{Rec}(\langle x\rangle^{\mathcal{C}}_p,\langle x\rangle^{\mathcal{S}}_p)=\langle x\rangle^{\mathcal{C}}_p\boxplus\langle x\rangle^{\mathcal{S}}_p=x$. Conventional $\textsf{Shr}(x)$ functionality in the state-of-the-art frameworks needs to exactly obtain the encrypted $x$ with massive amount of either computation-intensive \textsf{Rot} for a linear function or round-intensive \textsf{Mux} for the nonlinear function~\cite{rathee2020cryptflow2}. By carefully sharing specific intermediates in the process of computing adjacent functions, we efficiently eliminate all \textsf{Rot} and \textsf{Mux} in the cryptographic inference.

\begin{algorithm}[tb]
%\SetAlgoRefName{} % no count alg. num
\caption{Computation for \textsf{ReLU} and \textsf{Conv}, $\Pi_{\textsf{ReLU+Conv}}^{\textrm{ring},p}$}
\label{alg:reluconv}
\SetKwInput{KwInput}{Input}
\SetKwInput{KwOutput}{Output}
\DontPrintSemicolon
\KwInput{$\langle{\bm{a}}\rangle_p^{\mathcal{C}}$ from $\mathcal{C}$, and $\langle{\bm{a}}\rangle_p^{\mathcal{S}}$, $\textrm{\textbf{K}}\in\mathbb{Z}_{p}^{c_o\times{c_i}\times{f_h}\times{f_w}}$ fro- m $\mathcal{S}$ where $\bm{a}$ $\in{\mathbb{Z}_{p}^{c_i\times{h_i}\times{w_i}}}$.}
\KwOutput{$\mathcal{C}$ and $\mathcal{S}$ get $\langle{\bm{y}}\rangle_p^{\mathcal{C}}$ and $\langle{\bm{y}}\rangle_p^{\mathcal{S}}$, respectively, wh -ere $\bm{y}=\textsf{Conv}(\textsf{ReLU}(\bm{a}),\textrm{\textbf{K}})\in{\mathbb{Z}_{p}^{c_o\times{h_i'}\times{w_i'}}}$.}
\tcc{get shares of $\bar{\bm{a}}=\textsf{ReLU}(\bm{a})$}\tikzmk{A}
$\mathcal{S}$ gets $\langle{\hat{\bm{a}}}\rangle_2^{\mathcal{S}}$ $\scriptstyle\overset{\$}{\gets}$ $\mathbb{Z}_{2}^{c_i\times{h_i}\times{w_i}}$.\;
$\mathcal{C}$ interacts with $\mathcal{S}$ based on OT to obtain $\langle{\hat{\bm{a}}}\rangle_2^{\mathcal{C}}$ where $\textsf{Rec}($ $\langle{\hat{\bm{a}}}\rangle_2^{\mathcal{C}}, \langle{\hat{\bm{a}}}\rangle_2^{\mathcal{S}})=\hat{\bm{a}}=\textsf{DReLU}(\bm{a})\in{\mathbb{Z}_{2}^{c_i\times{h_i}\times{w_i}}}$.\;\tikzmk{B}\boxit{gray}
{\color[RGB]{28 28 28}$\mathcal{C}$ and $\mathcal{S}$ compute the $\textsf{Mux}(\bm{a},\hat{\bm{a}})=\bar{\bm{a}}=\textsf{ReLU}(\bm{a})$ to- gether, and get $\langle{\bar{\bm{a}}}\rangle_p^{\mathcal{C}}$ and $\langle{\bar{\bm{a}}}\rangle_p^{\mathcal{S}}$ where $\textsf{Rec}(\langle{\bar{\bm{a}}}\rangle_p^{\mathcal{C}},\langle{\bar{\bm{a}}}\rangle_p^{\mathcal{S}}$ $)=\bar{\bm{a}}\in{\mathbb{Z}_{p}^{c_i\times{h_i}\times{w_i}}}$.\;}
\tcc{get shares of $\bm{y}=\textsf{Conv}(\textsf{ReLU}(\bm{a}),\textrm{\textbf{K}})$}\tikzmk{A}
$\mathcal{C}$ encrypts $\langle{\bar{\bm{a}}}\rangle_p^{\mathcal{C}}$ as \textsf{Enc}$^{{\mathcal{C}}}({\langle{\bar{\bm{a}}}\rangle_p^{\mathcal{C}}})$ and sends it to $\mathcal{S}$.\;
$\mathcal{S}$ obtains \textsf{Enc}$^{\mathcal{C}}(\bar{\bm{a}})=$ \textsf{Enc}$^{{\mathcal{C}}}({\langle{\bar{\bm{a}}}\rangle_p^{\mathcal{C}}})$ $\oplus\langle{\bar{\bm{a}}}\rangle_p^{\mathcal{S}}$, and gets t- he $\textsf{Enc}^{\mathcal{C}}(\bm{y})$ based on \textsf{Enc}$^{\mathcal{C}}(\bar{\bm{a}})$ and $\textrm{\textbf{K}}$ through a seri- es of $\oplus$, $\otimes$ and \textsf{Rot}.\;
$\mathcal{S}$ samples $\langle{{\bm{y}}}\rangle_p^{\mathcal{S}}$ $\scriptstyle\overset{\$}{\gets}$ $\mathbb{Z}_{p}^{c_o\times{h_i'}\times{w_i'}}$.\;\tikzmk{B}\boxit{cyan}
{\color[RGB]{0,60,60}$\mathcal{S}$ computes the \textsf{Enc}$^{\mathcal{C}}(\langle{\bm{y}}\rangle_p^{\mathcal{C}})= \textsf{Enc}^{\mathcal{C}}(\bm{y})\ominus\langle{{\bm{y}}}\rangle_p^{\mathcal{S}}$, whi- ch is then sent to $\mathcal{C}$.\;
$\mathcal{C}$ obtains $\langle{\bm{y}}\rangle_p^{\mathcal{C}}=\textsf{Dec}^{\mathcal{C}}(\textsf{Enc}^{\mathcal{C}}(\langle{\bm{y}}\rangle_p^{\mathcal{C}}))$.}\;
\end{algorithm}
\subsection{Linear and Nonlinear Transformations}
Deevashwer Rathee et al.~\cite{rathee2020cryptflow2} have recently proposed efficient computation for linear and nonlinear functions $\Pi_{\textsf{ReLU+Conv}}^{\textrm{ring},p}$ as shown in Algorithm~\ref{alg:reluconv}, which contributes to high-performance neural network inference. Specifically, there are mainly two parts in $\Pi_{\textsf{ReLU+Conv}}^{\textrm{ring},p}$ where the first one is to get the shares of \textsf{ReLU} based on the shares of previous function (as shown in the gray block of Algorithm~\ref{alg:reluconv}), and the second one is to subsequently get the shares of \textsf{Conv} based on the shares of \textsf{ReLU} (as shown in the
blue block of Algorithm~\ref{alg:reluconv}). Here, the \textsf{ReLU} and \textsf{Conv} are analyzed together as they are always adjacent and serve as a repeatable module to form modern neural networks~\cite{simonyan2014very,he2016deep}. Among the state-of-the-art frameworks~\cite{liu2017oblivious,juvekar2018gazelle,mohassel2017secureml,riazi2019xonn,mishra2020delphi,rathee2020cryptflow2,huang2022cheetah,hussain2021coinn}, the above two parts are independently optimized such that the efficiency-unfriendly \textsf{Mux} and \textsf{Rot} are extensively involved to produce shares of one function, which serve as the input to compute subsequent function. While it is definitely logical to address the efficiency optimization in function wise, we show in the following section that our design enables efficient elimination of \textsf{Mux} and \textsf{Rot} by jointly computing the \textsf{ReLU} and \textsf{Conv}. Generally, specific intermediates of \textsf{ReLU} are utilized to directly compute \textsf{Conv} and vice versa. Such joint computation produces a new building block $\Pi_{\textsf{ReConv}}^{\textrm{ring},p}$ that combines the computation process of stacked functions and achieves more efficient privacy-preserving neural network inference.

\section{System Description}\label{sysdescrib}

\begin{figure}[!tbp]
\centering
\includegraphics[trim={7.8cm 3cm 7.3cm 1.2cm}, clip, scale=0.57]{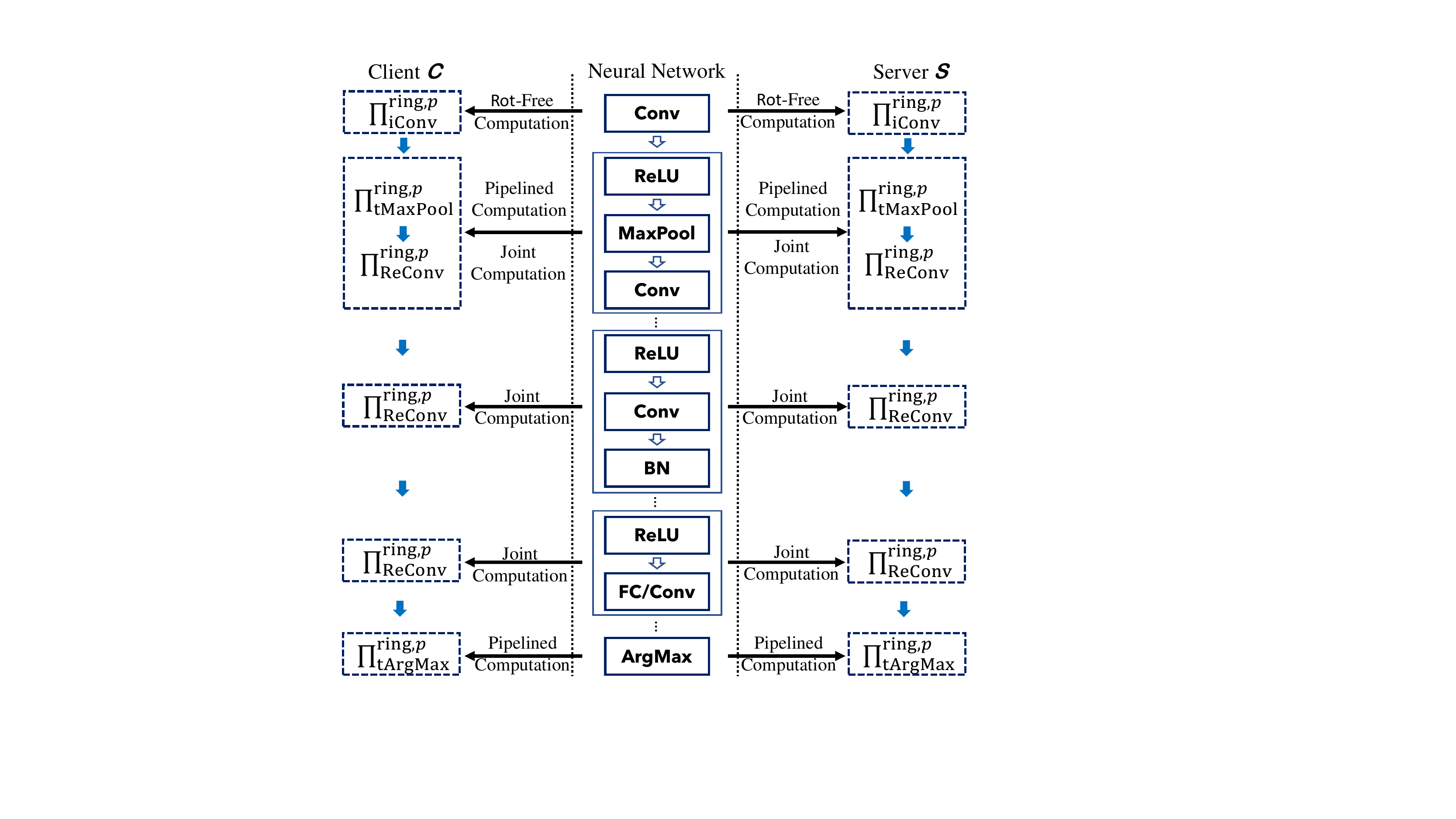}
%\vspace*{-0.25in}
\caption{Overview of the proposed framework.}
%\vspace*{-0.5in}
\label{fig:overview}
%\vspace*{-0.06in}
\end{figure}
\subsection{Overview}
\textcolor{black}{Figure~\ref{fig:overview} illustrates the overview of our proposed framework for cryptographic inference.} Specifically, a neural network contains a stack of linear and nonlinear functions. As for a CNN, we group these functions into six blocks as: 1) the \textsf{Conv} block for $\mathcal{C}$'s input (i.e., the first function in the CNN); 2) the $\textsf{ReLU}+\textsf{MaxPool}+\textsf{Conv}$ block; 3) the $\textsf{ReLU}+\textsf{Conv}+\textsf{BN}$ block; 4) the $\textsf{ReLU}+\textsf{Conv}$ block; 5) the $\textsf{ReLU}+\textsf{FC}$ block; and 6) the \textsf{ArgMax} block for network's output (i.e., the last function in the CNN). Such six blocks are able to construct many state-of-the-art CNN architectures such as \texttt{VGG}~\cite{simonyan2014very} and \texttt{ResNet}~\cite{he2016deep}. Compared with the function-wise optimization in existing schemes, which inevitably involves noticeable amount of expensive \textsf{Rot} and round-intensive \textsf{Mux}, we are motivated to jointly consider the computation process of functions in each block, aiming to efficiently eliminate the \textsf{Rot} and \textsf{Mux}, and thus to boost the performance of privacy-preserving inference.

As for the $\textsf{ReLU}+\textsf{Conv}$ block, we design the joint computation module $\Pi_{\textsf{ReConv}}^{\textrm{ring},p}$ (to be discussed in Sec.~\ref{sys:jointcom}), which serves as the core component in our framework. The $\textsf{ReLU}+\textsf{MaxPool}+\textsf{Conv}$ block is converted into the combination of $\Pi_{\textsf{ReConv}}^{\textrm{ring},p}$ and our optimized module for \textsf{MaxPool}, $\Pi_{\textsf{tMaxPool}}^{\textrm{ring},p}$ (to be explained in Sec.~\ref{further_opt}). For neural network inference, the \textsf{BN} in $\textsf{ReLU}+\textsf{Conv}+\textsf{BN}$ block is integrated into the \textsf{Conv}, and it enables us to compute such transformed block by $\Pi_{\textsf{ReConv}}^{\textrm{ring},p}$. The calculation towards \textsf{Conv} block for $\mathcal{C}$'s input and $\textsf{ReLU}+\textsf{FC}$ block works similarly with $\Pi_{\textsf{ReConv}}^{\textrm{ring},p}$. Finally, the \textsf{ArgMax} block for network's output is realized by our optimized module for \textsf{ArgMax}, $\Pi_{\textsf{tArgMax}}^{\textrm{ring},p}$ (to be marrated in Sec.~\ref{further_opt}). In the following, we first detail the construction of $\Pi_{\textsf{ReConv}}^{\textrm{ring},p}$ in Sec.~\ref{sys:jointcom} and then elaborate other optimizations in Sec.~\ref{further_opt}. Additionally, we follow the truncation strategy over $\mathbb{Z}_{p}$~\cite{rathee2020cryptflow2} to deal with the floating-point computation.

\subsection{The Joint Computation Module: $\Pi_{\textsf{ReConv}}^{\textrm{ring},p}$}\label{sys:jointcom}

For a lucid illustration, we begin with revisiting traditional calculation for the $\textsf{ReLU}+\textsf{Conv}$ block, where the \textit{PHE Triplet} is proposed in Sec. 3.2.1 for \textsf{Mux}-free \textsf{ReLU} and our \textit{Matrix Encoding} is described in Sec. 3.2.2 for \textsf{Rot}-free \textsf{Conv}.
These two components form our joint computation module $\Pi_{\textsf{ReConv}}^{\textrm{ring},p}$. Different from the compute-and-share process in state-of-the-art function-wise computation, which exhaustively obtains encrypted output of current function and then shares that result to serve as the input for next function, $\Pi_{\textsf{ReConv}}^{\textrm{ring},p}$ shares specifically formed intermediates to finish computing $\textsf{ReLU}+\textsf{Conv}$ and we refer this methodology as share-in-the-middle logic (SIM). SIM enables efficient elimination of \textsf{Rot} and \textsf{Mux}, which are expensive in the state-of-the-art solutions. The overall process of $\Pi_{\textsf{ReConv}}^{\textrm{ring},p}$ is summarized in
Sec. 3.2.3, which is followed by the corresponding complexity analysis in Sec. 3.2.4. The security of $\Pi_{\textsf{ReConv}}^{\textrm{ring},p}$ is
justified in Sec. 3.2.5.

\noindent\textbf{3.2.1 PHE Triplet Generation in \textsf{ReLU}.}
Recall in Algorithm~\ref{alg:reluconv} the traditional computation for $\textsf{ReLU}+\textsf{Conv}$, $\Pi_{\textsf{Relu+Conv}}^{\textrm{ring},p}$, where the input is the shares of previous function, $\langle{\bm{a}}\rangle_p^{\mathcal{C}}$ and $\langle{\bm{a}}\rangle_p^{\mathcal{S}}$. The conventional logic to compute \textsf{ReLU} includes the OT-based \textsf{DReLU} and subsequent \textsf{Mux} to obtain the output shares namely $\langle{\bar{\bm{a}}}\rangle_p^{\mathcal{C}}$ and $\langle{\bar{\bm{a}}}\rangle_p^{\mathcal{S}}$, which serve as the input of \textsf{Conv} (see the gray block in Algorithm~\ref{alg:reluconv}). Since almost one third of the communication rounds are consumed by \textsf{Mux} to obtain the output shares of \textsf{ReLU}~\cite{rathee2020cryptflow2}, we utilize the intermediates
from \textsf{DReLU} to directly enable the subsequent \textsf{Conv}. This design efficiently removes the \textsf{Mux} in the process of $\textsf{ReLU}+\textsf{Conv}$. Specifically, given the shares of $\textsf{DReLU}(\bm{a})$, $\langle{\hat{\bm{a}}}\rangle_2^{\mathcal{C}}$ and $\langle{\hat{\bm{a}}}\rangle_2^{\mathcal{S}}$, the \textsf{ReLU} of $\bm{a}$ is computed according to
\begin{align}
\label{eq1}&\textsf{ReLU}(\bm{a})=\textsf{DReLU}(\bm{a})\boxdot\bm{a}\\
\label{eq3}&=\underbrace{\{\langle{\hat{\bm{a}}}\rangle_2^{\mathcal{C}}\boxplus\langle{\hat{\bm{a}}}\rangle_2^{\mathcal{S}}\boxminus(\bm{2}\boxdot\langle{\hat{\bm{a}}}\rangle_2^{\mathcal{C}}\boxdot\langle{\hat{\bm{a}}}\rangle_2^{\mathcal{S}})\}}_{\circled{1}:\;\textrm{each}\;\textrm{operation}\;\textrm{is}\;\textrm{over}\;\mathbb{Z}_{p}^{c_i\times{h_i}\times{w_i}}}\boxdot(\langle{\bm{a}}\rangle_p^{\mathcal{C}}\boxplus\langle{\bm{a}}\rangle_p^{\mathcal{S}})\\
&=\underbrace{(\langle{\bm{a}}\rangle_p^{\mathcal{C}}\boxdot\langle{\hat{\bm{a}}}\rangle_2^{\mathcal{C}}\boxminus{\color{black}{\bm{r}^{\mathcal{C}}}})}_{\bm{h}_1}\boxplus\{\underbrace{\langle{\bm{a}}\rangle_p^{\mathcal{C}}\boxdot\{\bm{1}\boxminus(\bm{2}\boxdot\langle{\hat{\bm{a}}}\rangle_2^{\mathcal{C}})\}}_{\bm{h}_2}\boxdot\underbrace{{\color{black}\langle{\hat{\bm{a}}}\rangle_2^{\mathcal{S}}}}_{{\color{black}\bm{h}_3}}\}\boxplus\notag\\
\label{eq4}&\{\underbrace{{\color{black}\langle{\bm{a}}\rangle_p^{\mathcal{S}}\boxdot\{\bm{1}\boxminus(\bm{2}\boxdot\langle{\hat{\bm{a}}}\rangle_2^{\mathcal{S}})\}}}_{{\color{black}\bm{h}_4}}\boxdot\underbrace{\langle{\hat{\bm{a}}}\rangle_2^{\mathcal{C}}}_{{\color{black}\bm{h}_5}}\}\boxplus(\underbrace{\langle{\bm{a}}\rangle_p^{\mathcal{S}}\boxdot\langle{\hat{\bm{a}}}\rangle_2^{\mathcal{S}}}_{{\color{black}\bm{h}_6}})\boxplus\underbrace{\color{black}\bm{r}^{\mathcal{C}}}_{{\color{black}\bm{h}_7}}.
\end{align}

Here $``\boxdot"$, $``\boxplus"$ and $``\boxminus"$ are element-wise plaintext multiplication, addition and subtraction, respectively.
The item \circled{1} converts the shares of $\textsf{DReLU}(\bm{a})$ from $\mathbb{Z}_{2}^{c_i\times{h_i}\times{w_i}}$ to $\mathbb{Z}_p^{c_i\times{h_i}\times{w_i}}$. By rearranging and modifying the terms in Eq.~\eqref{eq3}, we finally get five to-be-added parts in Eq.~\eqref{eq4} to obtain \textsf{ReLU}$(\bm{a})$: 1) the $\mathcal{C}$-computed $\bm{h}_1$ where $\bm{r}^{\mathcal{C}}{\scriptstyle\overset{\$}{\gets}}\,\mathbb{Z}_p^{c_i\times{h_i}\times{w_i}}$; 2) the multiplication between $\mathcal{C}$-computed $\bm{h}_2$ and $\bm{h}_3$ namely $\langle{\hat{\bm{a}}}\rangle_2^{\mathcal{S}}$; 3) the multiplication between $\mathcal{S}$-computed $\bm{h}_4$ and $\bm{h}_5$ namely $\langle{\hat{\bm{a}}}\rangle_2^{\mathcal{C}}$; 4) the $\mathcal{S}$-computed $\bm{h}_6$; and 5) the $\bm{h}_7$ namely $\bm{r}^{\mathcal{C}}$. Given the input-independent nature of shares $\langle{\hat{\bm{a}}}\rangle_2^{\mathcal{S}}$, $\langle{{\bm{a}}}\rangle_p^{\mathcal{S}}$ and $\bm{r}^{\mathcal{C}}$ which are pregenerated by either $\mathcal{S}$ or $\mathcal{C}$ before the inference process, the $\bm{h}_3$, $\bm{h}_4$, $\bm{h}_6$ and $\bm{h}_7$ are pre-determined.
Meanwhile, $\mathcal{C}$ would encrypt its share of $\textsf{ReLU}(\bm{a})$ and send it to $\mathcal{S}$ for obtaining encrypted $\textsf{ReLU}(\bm{a})$ and thus enabling the subsequent HE-based \textsf{Conv}. By carefully encrypting the pre-determined terms, without introducing \textsf{Mux}, we are able to make $\mathcal{S}$ obtain the encrypted input for \textsf{Conv}.

Therefore, we define the PHE triplet:
\[\{\textsf{ct}^{\mathcal{S}}_{\color{black}\bm{h}_3}, \textsf{ct}^{\mathcal{S}}_{\color{black}\bm{h}_4}, \textsf{ct}^{\mathcal{C}}_{\color{black}\bm{h}_{7}}\}\]
where the first two terms are generated by $\mathcal{S}$ and sent to $\mathcal{C}$ while the last term is generated by $\mathcal{C}$ and sent to $\mathcal{S}$. Meanwhile, it is worth pointing out that each component in the above triplet is non-interactively formed and sent to either $\mathcal{C}$ or $\mathcal{S}$ \textit{offline}. As such, $\mathcal{S}$ precomputes
\begin{equation}\label{eq:ct_ch8}
\textsf{ct}_{\bm{h}_8}^{\mathcal{C}}=\bm{h}_6\oplus{\textsf{ct}^{\mathcal{C}}_{\color{black}\bm{h}_{7}}}
\end{equation}
while $\mathcal{C}$ obtains
\begin{equation}\label{ct_sh9}
\textsf{ct}_{\bm{h}_9}^{\mathcal{S}}=\bm{h}_1\oplus{(\bm{h}_2\otimes\textsf{ct}^{\mathcal{S}}_{\color{black}\bm{h}_3})\oplus({\textsf{ct}^{\mathcal{S}}_{\color{black}\bm{h}_4}\otimes{\bm{h}_5}})}
\end{equation}
and sends it to $\mathcal{S}$ right after the \textsf{DReLU} computation. $\mathcal{S}$ then conducts the decryption as
\begin{equation}\label{h9}
\bm{h}_9=\textsf{Dec}^{\mathcal{S}}(\textsf{ct}_{\bm{h}_9}^{\mathcal{S}})=\bm{h}_1\boxplus{(\bm{h}_2\boxdot{\color{black}\bm{h}_3})\boxplus({{\color{black}\bm{h}_4}\boxdot{\bm{h}_5}})}.
\end{equation}

It is obvious that $\textsf{ReLU}(\bm{a})=\bm{h}_8\boxplus\bm{h}_9$. Thus $\mathcal{S}$ obtains the $\mathcal{C}$-encrypted $\textsf{ReLU}(\bm{a})$ as
\begin{equation}\label{ct_cabar}
\textsf{ct}_{\bar{\bm{a}}}^{\mathcal{C}}=\textsf{ct}_{\bm{h}_8}^{\mathcal{C}}\oplus\bm{h}_9,
\end{equation}
which is utilized to compute \textsf{Conv} to be elaborated next. Note that the proposed PHE triplet enables us to eliminate the \textsf{Mux} and directly get the encrypted \textsf{ReLU} at $\mathcal{S}$ for computing the subsequent \textsf{Conv}, with only half a round after the \textsf{DReLU}. This helps to save all computation and communication cost for calling \textsf{Mux}.

\noindent\textbf{3.2.2 Matrix Encoding for \textsf{Conv}.}
Based on the encrypted $\textsf{ReLU}$ $\textsf{ct}_{\bar{\bm{a}}}^{\mathcal{C}}$ in Eq.~\eqref{ct_cabar}, $\mathcal{S}$ is supposed to utilize its kernel $\textrm{\textbf{K}}$ to compute the \textsf{Conv} (see the blue block in Algorithm~\ref{alg:reluconv}). In other words, $\mathcal{S}$ should obtain Enc$^{\mathcal{C}}(\bm{y})=\textsf{Conv}(\textsf{ct}_{\bar{\bm{a}}}^{\mathcal{C}},\textrm{\textbf{K}})$ and share it with $\mathcal{C}$. The state-of-the-art frameworks extensively call the PHE operations to get the encrypted output of \textsf{Conv}, which is then shared between $\mathcal{C}$ and $\mathcal{S}$ to compute the subsequent function. However, it inevitably involves a series of \textsf{Rot} due to the needed summing process~\cite{juvekar2018gazelle}. We totally relieve the need for \textsf{Rot} via a carefully designed matrix encoding.

Our main idea lays in the observation that the dot product between a matrix and a vector in Eq.~\eqref{yj} can be viewed as
the linear combination of all columns in that matrix. As such, if the $\textsf{ReLU}(\bm{a})$ and kernel \textbf{K} were respectively such matrix and vector, we could construct $\textsf{ct}_{\bar{\bm{a}}}^{\mathcal{C}}$ as a set of ciphertext each of which encrypts one column in $\textsf{ReLU}(\bm{a})$, and the \textsf{Conv} output, Enc$^{\mathcal{C}}(\bm{y})$, could be  obtained by $\mathcal{S}$ via only HE multiplication and addition. In the following, we describe in detail the feasibility of above conjecture for \textsf{Rot}-free \textsf{Conv}, and Sec. 3.2.4 shows the efficiency advantages of our design compared with the state-of-the-art solution.

Specifically, we first transform the \textsf{Conv} $\bm{y}=\textsf{Conv}(\bar{\bm{a}},\textbf{\textrm{K}})$ $\in\mathcal{R}^{c_o\times h_i'\times{w_i'}}$ into corresponding dot product $\tilde{\bm{y}}=\textsf{Dot}(\tilde{\bm{a}},\tilde{\textbf{\textrm{K}}})$ $\in\mathcal{R}^{h_i'w_i'\times{c_o}}$ based on the im2col operator~\cite{jia2014caffe}. Recall that $\bar{\bm{a}}\in\mathcal{R}^{c_i\times h_i\times w_i}$ and ${\textbf{K}}\in\mathcal{R}^{c_o\times c_i\times f_h\times f_w}$, they are then respectively converted into $\tilde{\bm{a}}\in\mathcal{R}^{h_i'w_i'\times{c_if_hf_w}}$ and $\widetilde{\textbf{K}}\in\mathcal{R}^{c_if_hf_w\times{c_o}}$ such that $\tilde{\bm{y}}_{\alpha,\beta}=\bm{y}_{\beta,\gamma,\eta}$ where $\alpha\in[h_i'w_i']$, $\beta\in[c_o]$, $\gamma=\left \lfloor \frac{\alpha }{w_i'} \right \rfloor$, $h_i'=\left \lceil \frac{h_i}{s} \right \rceil$, $w_i'=\left \lceil \frac{w_i}{s} \right \rceil$, $\eta=\alpha$ mod $w_i'$. In other words, each $c_if_hf_w$ elements along the first dimension of \textbf{K} forms one column in $\widetilde{\textbf{K}}$ while each $c_if_hf_w$ values in $\bar{\bm{a}}$ that are weighted and summed for one number of $\bm{y}$ forms a row in $\tilde{\bm{a}}$.
Thereafter, we denote the mapping from ${\bm{y}}$ to $\tilde{\bm{y}}$ as $\phi: \bm{y}\mapsto{\tilde{\bm{y}}}$.
Note that the conversion towards $\bar{\bm{a}}$ is completed during the \textsf{ReLU} process which is computed in element wise and is independent with element locations. Meanwhile the conversion towards \textbf{K} is easily performed by $\mathcal{S}$ since it's in plaintext. We also summarize the overall procedure in Sec. 3.2.3.

Recall that we are supposed to construct $\textsf{ct}_{\bar{\bm{a}}}^{\mathcal{C}}$
such that the computation for Enc$^{\mathcal{C}}(\bm{y})$ doesn't involve any \textsf{Rot}. Given $\tilde{\bm{a}}$, a matrix encoding is proposed which produces a set of ciphertext and enables \textsf{Rot}-free \textsf{Conv}. Concretely, we define the encoding mapping $\iota: \mathcal{R}^{h_i'w_i'\times{c_if_hf_w}}\mapsto\mathcal{R}^{d\times{n}}$ by
\[
\iota: \tilde{\bm{a}}\mapsto{\textbf{\textrm A}}
\]
where $\textbf{A}_{j,\zeta}=\tilde{\bm{a}}_{\tau,\lambda}$, $\lambda=j\xi+\left \lfloor \frac{\zeta}{h_i'w_i'} \right \rfloor$, $j\in[d]$, $\zeta\in[n]$, $\xi=\left \lfloor \frac{n}{h_i'w_i'} \right \rfloor$, $d=\left \lceil \frac{c_if_hf_w}{\xi} \right \rceil$, and $\tau=\zeta$ mod $h_i'w_i'$.
In this way, $\iota$ maps $\xi$ columns of $\tilde{\bm{a}}$ into one row of $\textbf{\textrm A}$. Meanwhile, $\forall\beta\in[c_o]$, we have $\tilde{\bm{y}}_{:,\beta}=\sum_{\lambda}\tilde{\bm{a}}_{:,\lambda}\widetilde{\textbf{\textrm{K}}}_{\lambda,\beta}$, which means that the $\beta$-th output channel after \textsf{Conv} is the summation of all weighted columns in $\tilde{\bm{a}}$ namely the $\lambda$-th cloumn $\tilde{\bm{a}}_{:,\lambda}$ is multiplied with $\widetilde{\textbf{\textrm{K}}}_{\lambda,\beta}$.
Therefore, $\forall\beta$, $\sum_{j}(\textbf{A}_{j,:}\boxdot\overline{\textbf{K}}_{j,\beta})\in\mathcal{R}^n$ contains $\xi$ finally summed columns derived from $\{\tilde{\bm{a}}_{:,\lambda}\tilde{\textbf{\textrm{K}}}_{\lambda,\beta}\}_{\lambda}$ where $\overline{\textbf{K}}_{j,\beta}\in\mathcal{R}^n=\{\widetilde{\textbf{\textrm{K}}}_{j\xi,\beta}\}^{h_i'w_i'}|\cdots|\{\widetilde{\textbf{\textrm{K}}}_{(j+1)\xi-1,\beta}\}^{h_i'w_i'}|\{0\}^{n-h_i'w_i'\xi}$ and note that $\widetilde{\textbf{\textrm{K}}}_{\varphi,\beta}=0$ if $\varphi\geq{c_if_hf_w}$.

Based on the element-wise operation of $\sum_{j}(\textbf{A}_{j,:}\boxdot\overline{\textbf{K}}_{j,\beta})$, we are ready to obtain a \textsf{Rot}-free \textsf{Conv}.
In particular, each $\textbf{A}_{j,:}$ is encrypted into the ciphertext $\textsf{ct}_{\textbf{A}_{j,:}}^{\mathcal{C}}$, then $\forall\beta$, we obtain a $\textsf{ct}_{\textbf{B}_{\beta}}^{\mathcal{C}}=\oplus_{j}(\textsf{ct}_{\textbf{A}_{j,:}}^{\mathcal{C}}\odot\overline{\textbf{K}}_{j,\beta})$ which encrypts $\xi$ finally summed columns derived from $\{\tilde{\bm{a}}_{:,\lambda}\tilde{\textbf{\textrm{K}}}_{\lambda,\beta}\}_{\lambda}$. Here $\textbf{B}_{\beta}\in\mathcal{R}^n=\sum_{j}({\textbf{A}_{j,:}}\boxdot\overline{\textbf{K}}_{j,\beta})$. Considering that $\textbf{B}_{\beta}$ contains all partial sums for $\tilde{\bm{y}}_{:,\beta}$ and $\{\textsf{ct}^{\mathcal{C}}_{\textbf{B}_{\beta}}\}_{\beta}$ are obtained by $\mathcal{S}$ without \textsf{Rot}, $\mathcal{S}$ then directly shares $\{\textsf{ct}^{\mathcal{C}}_{\textbf{B}_{\beta}}\}_{\beta}$ by $\{\textsf{ct}^{\mathcal{C}}_{\textbf{C}_{\beta}}=\textsf{ct}^{\mathcal{C}}_{\textbf{B}_{\beta}}\ominus{\bm{r}^{\mathcal{S}}_{\beta}}\}_{\beta}$ where $\textbf{C}_{\beta}\in\mathcal{R}^n=\textbf{B}_{\beta}\boxminus\bm{r}^{\mathcal{S}}_{\beta}$ and $\bm{r}^{\mathcal{S}}_{\beta}{\scriptstyle\overset{\$}{\gets}}\,\mathbb{Z}_p^{n}$. Then $\mathcal{S}$ sends $\{\textsf{ct}^{\mathcal{C}}_{\textbf{C}_{\beta}}\}_{\beta}$ to $\mathcal{C}$, which decrypts $\{\textsf{ct}^{\mathcal{C}}_{\textbf{C}_{\beta}}\}_{\beta}$ into $\{\textsf{Dec}^{\mathcal{C}}(\textsf{ct}^{\mathcal{C}}_{\textbf{C}_{\beta}})=\textbf{C}_{\beta}\}_{\beta}$ and gets
\begin{equation}\label{eq:ycp}
\left \langle \bm{y} \right \rangle_p^{\mathcal{C}}=\phi^{-1}(\left \langle \tilde{\bm{y}}_{:,0} \right \rangle_p^{\mathcal{C}}|\left \langle \tilde{\bm{y}}_{:,1} \right \rangle_p^{\mathcal{C}}|\cdots|\left \langle \tilde{\bm{y}}_{:,c_o-1} \right \rangle_p^{\mathcal{C}})
\end{equation}
where $\left \langle \tilde{\bm{y}}_{:,\beta} \right \rangle_p^{\mathcal{C}}=\{\boxplus_{\chi=0}^{\xi-1}\textbf{C}_{\beta,\{h_i'w_i'\chi:h_i'w_i'(\chi+1)\}}\}^{\top}$.
Meanwhile, $\mathcal{S}$ computes
\begin{equation}\label{eq:y_sp}
\left \langle \bm{y} \right \rangle_p^{\mathcal{S}}=\phi^{-1}(\left \langle \tilde{\bm{y}}_{:,0} \right \rangle_p^{\mathcal{S}}|\left \langle \tilde{\bm{y}}_{:,1} \right \rangle_p^{\mathcal{S}}|\cdots|\left \langle \tilde{\bm{y}}_{:,c_o-1} \right \rangle_p^{\mathcal{S}})
\end{equation}
where $\left \langle \tilde{\bm{y}}_{:,\beta} \right \rangle_p^{\mathcal{S}}=\{\boxplus_{\chi=0}^{\xi-1}\bm{r}^{\mathcal{S}}_{\beta,\{h_i'w_i'\chi:h_i'w_i'(\chi+1)\}}\}^{\top}$.
Here $\mathcal{S}$ is able to get $\left \langle \bm{y} \right \rangle_p^{\mathcal{S}}$ offline.
As such, $\mathcal{C}$ and $\mathcal{S}$ respectively obtain their shares of $\textsf{Conv}$, which act as the input of subsequent function.

\begin{algorithm}[tb]
%\SetAlgoRefName{} % no count alg. num
\caption{Joint computation of \textsf{ReLU} and \textsf{Conv}, $\Pi_{\textsf{ReConv}}^{\textrm{ring},p}$}
\label{alg:reconv}
\SetKwInput{KwInput}{Input}
\SetKwInput{KwOutput}{Output}
\DontPrintSemicolon
\KwInput{$\langle{\bm{a}}\rangle_p^{\mathcal{C}}$ from $\mathcal{C}$, and $\langle{\bm{a}}\rangle_p^{\mathcal{S}}$, $\textrm{\textbf{K}}\in\mathbb{Z}_{p}^{c_o\times{c_i}\times{f_h}\times{f_w}}$ fr- om $\mathcal{S}$ where $\bm{a}$ $\in{\mathbb{Z}_{p}^{c_i\times{h_i}\times{w_i}}}$.}
\KwOutput{$\mathcal{C}$ and $\mathcal{S}$ get $\langle{\bm{y}}\rangle_p^{\mathcal{C}}$ and $\langle{\bm{y}}\rangle_p^{\mathcal{S}}$, respectively, w- here $\bm{y}=\textsf{Conv}(\textsf{ReLU}(\bm{a}),\textrm{\textbf{K}})\in{\mathbb{Z}_{p}^{c_o\times{h_i'}\times{w_i'}}}$.}
\tcc{\underline{\textbf{offline computation}}}
$\mathcal{C}$ gets $\bm{r}^{\mathcal{C}}$ $\scriptstyle\overset{\$}{\gets}\,\mathbb{Z}_{p}^{c_i\times{h_i}\times{w_i}}$, maps $\bm{h}_7$ to ${\textbf{H}}$ based on $\iota$ whi- ch has the same structure as \textbf{A} (mapped from $\bar{\bm{a}}$), an- d encrypts ${\textbf{H}}$ as $\{\textsf{ct}_{\textbf{H}_{j,:}}^{\mathcal{C}}\}_{j\in[d]}$ which are sent to $\mathcal{S}$.\;
$\mathcal{S}$ maps \textbf{K} to $\overline{\textbf{K}}$, generates $\langle{\hat{\bm{a}}}\rangle_2^{\mathcal{S}}$ $\scriptstyle\overset{\$}{\gets}\,\mathbb{Z}_{2}^{c_i\times{h_i}\times{w_i}}$ and $\{\bm{r}^{\mathcal{S}}_{\beta}$ ${\scriptstyle\overset{\$}{\gets}}\,\mathbb{Z}_p^{n}\}_{\beta}$, maps $\bm{h}_6$ to  $\widehat{\textbf{H}}$ based on $\iota$ which has the sa- me structure as \textbf{A}, maps $\bm{h}_3$ and $\bm{h}_4$ to $\overline{\textbf{H}}$ and $\widetilde{\textbf{H}}$, res- pectively, where $\psi: \bm{h}_3\mapsto$ $\overline{\textbf{H}}$. Here $\overline{\textbf{H}}_{\nu,:}$ places each $n$ elements of $\bm{h}_3$ in a row where $\nu\in[\sigma=\left \lceil \frac{c_ih_iw_i}{n} \right \rceil]$,
and $\widetilde{\textbf{H}}$ has the same structure as $\overline{\textbf{H}}$. $\mathcal{S}$ also encrypts $\overline{\textbf{H}}$ and $\widetilde{\textbf{H}}$ to  $\{\textsf{ct}_{\overline{\textbf{H}}_{\nu,:}}^{\mathcal{S}}\}_{\nu}$ and $\{\textsf{ct}_{\widetilde{\textbf{H}}_{\nu,:}}^{\mathcal{S}}\}_{\nu}$ which are then s- ent to $\mathcal{C}$, obtains the $\left \langle \bm{y} \right \rangle_p^{\mathcal{S}}$ by Eq.~\ref{eq:y_sp}, and then comp- utes $\{\textsf{ct}_{\widecheck{\textbf{H}}_{j,:}}^{\mathcal{C}}=\widehat{\textbf{H}}_{j,:}\oplus\textsf{ct}_{{\textbf{H}}_{j,:}}^{\mathcal{C}}\}_j$ based on Eq.~\ref{eq:ct_ch8}.\;
\tcc{\underline{\textbf{$\mathcal{C}$'s share of} $\bm{y}=\textsf{Conv}(\textsf{ReLU}(\bm{a}),\textrm{\textbf{K}})$}}
$\mathcal{C}$ interacts with $\mathcal{S}$ based on OT to obtain $\langle{\hat{\bm{a}}}\rangle_2^{\mathcal{C}}$ where $\textsf{Rec}($ $\langle{\hat{\bm{a}}}\rangle_2^{\mathcal{C}}, \langle{\hat{\bm{a}}}\rangle_2^{\mathcal{S}})=\hat{\bm{a}}=\textsf{DReLU}(\bm{a})\in{\mathbb{Z}_{2}^{c_i\times{h_i}\times{w_i}}}$.\;
$\mathcal{C}$ maps $\bm{h}_1$, $\bm{h}_2$ and $\bm{h}_5$ to $\breve{\textbf{H}}$, $\ddot{\textbf{H}}$ and $\dot{\textbf{H}}$ based on $\psi$, re- spectively, which all have the same structures as $\overline{\textbf{H}}$, a -nd then computes $\{\textsf{ct}_{\acute{\textbf{H}}_{\nu,:}}^{\mathcal{S}}$= $\breve{\textbf{H}}_{\nu,:}\oplus(\ddot{\textbf{H}}_{\nu,:}\otimes{\textsf{ct}_{\overline{\textbf{H}}_{\nu,:}}^{\mathcal{S}}})\oplus($ $\textsf{ct}_{\widetilde{\textbf{H}}_{\nu,:}}^{\mathcal{S}}\otimes\dot{\textbf{H}}_{\nu,:})\}_{\nu}$ based on Eq.~\ref{ct_sh9}, which are sent to $\mathcal{S}$.\;
$\mathcal{S}$ decrypts $\{\textsf{ct}_{\acute{\textbf{H}}_{\nu,:}}^{\mathcal{S}}\}_{\nu}$ to $\{\acute{\textbf{H}}_{\nu,:}=\textsf{Dec}^{\mathcal{S}}(\textsf{ct}_{\acute{\textbf{H}}_{\nu,:}}^{\mathcal{S}})\}_{\nu}$ based on Eq.~\ref{h9}, gets $\bm{h}_9=\psi^{-1}(\acute{\textbf{H}})$, maps $\bm{h}_9$ to $\grave{\textbf{H}}$ based on $\iota$, computes $\{\textsf{ct}_{\textbf{A}_{j,:}}^{\mathcal{C}}$= $\textsf{ct}_{\widecheck{\textbf{H}}_{j,:}}^{\mathcal{C}}\oplus\grave{\textbf{H}}_{j,:}\}_j$ based on Eq.~\ref{ct_cabar}, ob -tains $\{\textsf{ct}^{\mathcal{C}}_{\textbf{B}_{\beta}}\}_{\beta\in[c_o]}$ and gets $\{\textsf{ct}^{\mathcal{C}}_{\textbf{C}_{\beta}}\}_{\beta}$ which are sent t -o $\mathcal{C}$.\;
$\mathcal{C}$ decrypts $\{\textsf{ct}^{\mathcal{C}}_{\textbf{C}_{\beta}}\}_{\beta}$ to $\{\textbf{C}_{\beta}\}_{\beta}$, and gets $\left \langle \bm{y} \right \rangle_p^{\mathcal{C}}$ based on Eq.~\ref{eq:ycp}.
\end{algorithm}
\noindent\textbf{3.2.3 Putting Things Together.}
With the proposed PHE triplet in \textsf{ReLU} and the matrix encoding for \textsf{Conv}, the \textsf{Mux} in \textsf{ReLU} and the \textsf{Rot} in \textsf{Conv} are totally eliminated, which forms our building block $\Pi_{\textsf{ReConv}}^{\textrm{ring},p}$ to compute consecutive \textsf{ReLU} and \textsf{Conv}. The complexity analysis to be elaborated in Sec. 3.2.4 demonstrates the numerical advantages of $\Pi_{\textsf{ReConv}}^{\textrm{ring},p}$ compared with the state-of-the-art function-wise design~\cite{rathee2020cryptflow2}. Algorithm~\ref{alg:reconv}
summaries the joint computation process.
Specifically, we have an offline phase for precomputation and an online phase for getting $\langle{\bm{y}}\rangle_p^{\mathcal{C}}$ based on the shares of \textsf{DReLU}. Since the dimension of $\bar{\bm{a}}$ is the same as the ones of seven items in Eq.~\eqref{eq4}, the structures of $\bm{h}_1$ to $\bm{h}_7$ are correspondingly tuned to transform $\textsf{ct}_{\bar{\bm{a}}}^{\mathcal{C}}$ to $\{\textsf{ct}_{\textbf{\textrm A}_{j,:}}^{\mathcal{C}}\}_j$.

Concretely, the structures of $\bm{h}_6$ and $\bm{h}_7$ are mapped to the one of \textbf{A}. Meanwhile, as $\bm{h}_9$ is homomorphically obtained by $\mathcal{C}$ based on Eq.~\ref{ct_sh9} and is then decrypted by $\mathcal{S}$ based on Eq.~\ref{h9}, the structures of $\bm{h}_1$ to $\bm{h}_5$ can be as tight as possible (see the mapping $\psi$ in Algorithm~\ref{alg:reconv}) to enable minimal calls of PHE operations at both $\mathcal{S}$ and $\mathcal{C}$. Furthermore, the mapped $\bm{h}_9$ at $\mathcal{S}$ has the same structure as \textbf{A}, which enables $\mathcal{S}$ to compute $\{\textsf{ct}_{\textbf{\textrm A}_{j,:}}^{\mathcal{C}}\}_j$ based on Eq.~\ref{ct_cabar}. After that, $\mathcal{S}$ gets $\{\textsf{ct}^{\mathcal{C}}_{\textbf{C}_{\beta}}\}_{\beta}$ which are later decrypted by $\mathcal{C}$ as $\{\textbf{C}_{\beta}\}_{\beta}$. $\mathcal{C}$ finally gets $\left \langle \bm{y} \right \rangle_p^{\mathcal{C}}$ based on Eq.~\ref{eq:ycp}. The offline computation mainly involves the generation of our PHE triplet whose non-interactive and input-independent nature relieves the necessary synchronization required in other state-of-the-art offline processes~\cite{liu2017oblivious,mishra2020delphi}. After $\Pi_{\textsf{ReConv}}^{\textrm{ring},p}$, the shares of \textsf{Conv} $\bm{y}$ act as the input of subsequent function.
Note that the bias $\bm{b}\in\mathcal{R}^{c_o}$ in the \textsf{Conv} is combined with $\{\bm{r}^{\mathcal{S}}_{\beta}\}_{\beta}$ such that
\[
\left \langle \tilde{\bm{y}}_{:,\beta} \right \rangle_p^{\mathcal{S}}=\{\boxplus_{\chi=0}^{\xi-1}\bm{r}^{\mathcal{S}}_{\beta,(h_i'w_i'\chi:h_i'w_i'(\chi+1))}\}^{\top}\boxplus\bm{b}_{\beta}.
\]

Furthermore, our joint computation block is easily adapted for
the $\textsf{ReLU}+\textsf{FC}$ block.
Concretely, the $\bar{\bm{a}}=\textsf{ReLU}(\bm{a})$ becomes an $n_i$-size vector and the mapping $\iota$ becomes
\[
\iota: \bar{\bm{a}}\mapsto{\textbf{A}}
\]
where $\textbf{A}\in\mathcal{R}^{{n}}=\{\bar{\bm{a}}\}^{\xi}|\{0\}^{n-n_i\xi}$. The weight matrix $\textbf{W}\in\mathcal{R}^{n_o\times{n_i}}$ is transformed to $\bar{\textbf{W}}\in\mathcal{R}^{d_1\times{n}}$ where $d_1\in{[\left \lceil \frac{n_o}{\xi} \right \rceil]}$, $\xi=\left \lfloor \frac{n}{n_i} \right \rfloor$ and $\bar{\textbf{W}}_{j,:}=\textbf{W}_{j\xi,:}|\cdots|\textbf{W}_{(j+1)\xi-1,:}|\{0\}^{n-n_i\xi}$, $j\in[d_1]$, $\textbf{W}_{\phi,:}=\{0\}^{n_i}$ if $\phi\geq{n_o}$. In this way, the $\textsf{ct}^{\mathcal{C}}_{\bar{\bm{a}}}$ obtained by $\mathcal{S}$, based on Eq.~\ref{ct_cabar}, turns out to be $\textsf{ct}^{\mathcal{C}}_{\textbf{A}}$ that encrypts $\xi$ copies of $\bar{\bm{a}}$ and
the ciphertext $\textsf{ct}^{\mathcal{C}}_{\textbf{B}_j}=\textsf{ct}^{\mathcal{C}}_{\textbf{A}}\otimes{\bar{\textbf{W}}_{j,:}}$ contains all partial sums of $j\xi$-th to $(j\xi+\xi-1)$-th components in $\bm{y}=\textsf{Dot}(\textbf{W},\bar{\bm{a}})\in\mathcal{R}^{n_o}$. Without introducing the \textsf{Rot}, $\mathcal{S}$ shares $\{\textsf{ct}^{\mathcal{C}}_{\textbf{B}_j}\}_{j}$ by $\{\textsf{ct}^{\mathcal{C}}_{\textbf{C}_{j}}=\textsf{ct}^{\mathcal{C}}_{\textbf{B}_{j}}\ominus{\bm{r}^{\mathcal{S}}_{j}}\}_{j}$ and sends $\{\textsf{ct}^{\mathcal{C}}_{\textbf{C}_{j}}\}_{j}$ to $\mathcal{C}$, which decrypts $\{\textsf{ct}^{\mathcal{C}}_{\textbf{C}_{j}}\}_{j}$ into $\{\textsf{Dec}^{\mathcal{C}}(\textsf{ct}^{\mathcal{C}}_{\textbf{C}_{j}})=\textbf{C}_{j}\}_{j}$ and gets
\begin{equation}
\left \langle \bm{y} \right \rangle_p^{\mathcal{C}}=\left \langle {\bm{y}}_{0} \right \rangle_p^{\mathcal{C}}|\left \langle {\bm{y}}_{1} \right \rangle_p^{\mathcal{C}}|\cdots|\left \langle {\bm{y}}_{n_o-1} \right \rangle_p^{\mathcal{C}}
\end{equation}
where $\left \langle {\bm{y}}_{\beta\in[n_o]} \right \rangle_p^{\mathcal{C}}=\boxplus_{\chi=0}^{n_i-1}\textbf{C}_{\tau,n_i\lambda+\chi}$, $\tau=\left \lfloor \frac{\beta}{\xi} \right \rfloor$, $\lambda=\beta\;\textrm{mod}\;\xi$.
Meanwhile, $\mathcal{S}$ computes
\begin{equation}
\left \langle \bm{y} \right \rangle_p^{\mathcal{S}}=\left \langle {\bm{y}}_{0} \right \rangle_p^{\mathcal{S}}|\left \langle {\bm{y}}_{1} \right \rangle_p^{\mathcal{S}}|\cdots|\left \langle {\bm{y}}_{n_o-1} \right \rangle_p^{\mathcal{S}}
\end{equation}
in offline and here $\left \langle {\bm{y}}_{\beta} \right \rangle_p^{\mathcal{S}}=\boxplus_{\chi=0}^{n_i-1}\bm{r}^{\mathcal{S}}_{\tau,n_i\lambda+\chi}$. Similar with \textsf{Conv}, the bias $\bm{b}\in\mathcal{R}^{n_o}$ is combined with $\{\bm{r}^{\mathcal{S}}_j\}_j$ such that
\[
\left \langle {\bm{y}}_{\beta} \right \rangle_p^{\mathcal{S}}=\{\boxplus_{\chi=0}^{n_i-1}\bm{r}^{\mathcal{S}}_{\tau,n_i\lambda+\chi}\}\boxplus\bm{b}_{\beta}.
\]

\begin{table*}[!tb]
\centering
\scriptsize
\caption{Complexity comparison to the state-of-the-art framework.}\label{complexity}
\vspace*{-0.15in}
\begin{tabular}{||c|c c c c c | c c c c c ||}
\hline
\multirow{2}{*}{Framework} & \multicolumn{5}{c|}{Before \textsf{Conv}} &\multicolumn{5}{c||}{For \textsf{Conv}}\\
\cline{2-11}
& \# Comm. Round & \# \textsf{Enc} & \# \textsf{Mult} & \# \textsf{Dec}& \# \textsf{Add}  & \# \textsf{Rot} & \# \textsf{Mult} & \# \textsf{Dec} & \# \textsf{Add} & \# Chiphertext\\
\hline
~\cite{rathee2020cryptflow2}& 4.5& $\geq\sigma$ & 0 & 0 &$\geq\sigma$  & $\geq(f_hf_w-1)\sigma+c_o-\left \lceil \frac{c_o}{\left \lfloor \frac{n}{h_iw_i} \right \rfloor} \right \rceil$& $\approx{dc_o}$ & $\left \lceil \frac{c_o}{\left \lfloor \frac{n}{h_iw_i} \right \rfloor} \right \rceil$  &$\approx{dc_o}$ & $\left \lceil \frac{c_o}{\left \lfloor \frac{n}{h_iw_i} \right \rfloor} \right \rceil$ \\
\hline
Ours& 0.5& 0& $2\sigma$ & $\sigma$ & ${2\sigma+d}$  &0 & $dc_o$& $c_o$ & $dc_o$ & $c_o$\\
\hline
\end{tabular}
\end{table*}
\noindent\textbf{3.2.4 Complexity Analysis.}
We now give the complexity analysis\footnote{We mainly analyze the computation and communication complexity in the online phase when $\mathcal{C}$ feeds her private input to compute the output of neural network. Similar analysis is applied for the offline computation.} for the proposed
$\Pi_{\textsf{ReConv}}^{\textrm{ring},p}$, and demonstrate its numerical advantages over the function-wise computation for consecutive \textsf{ReLU} and \textsf{Conv} in the state-of-the-art framework~\cite{rathee2020cryptflow2}\footnote{Note that~\cite{rathee2020cryptflow2} gives optimal complexity under the basic PHE and OT primitives with full security and intact neural networks among the state-of-the-art frameworks~\cite{liu2017oblivious,juvekar2018gazelle,mishra2020delphi,rathee2020cryptflow2}.} as shown in Algorithm~\ref{alg:reluconv}. Specifically,~\cite{rathee2020cryptflow2} introduces the \textsf{Mux} to get the shares of $\textsf{ReLU}(\bm{a})$ after the \textsf{DReLU}, see the line 2 in Algorithm~\ref{alg:reluconv}, which involves four communication rounds\footnote{Two rounds in parallel.} with $2c_ih_iw_i(\kappa+2\left \lceil \log{p} \right \rceil)$ bits where the $\kappa$ is the security parameter. After that $\mathcal{C}$ encrypts her share of \textsf{ReLU} to more than $\sigma$ ciphertext which are sent to $\mathcal{S}$. $\mathcal{S}$ then conducts more than $\sigma$ PHE addition $(\textsf{Add})$ to get encrypted $\textsf{ReLU}$, which serves as the input of \textsf{Conv}. In our protocol, $\mathcal{C}$ sends $\sigma$ ciphertext to $\mathcal{C}$ after $2\sigma$ PHE multiplication $(\textsf{Mult})$ and $2\sigma$ \textsf{Add}, as illustrated in line 4 of Algorithm~\ref{alg:reconv}. $\mathcal{C}$ subsequently conducts $\sigma$ PHE decryption $(\textsf{Dec})$ and $d$ \textsf{Add} to get $d$ ciphertext that act as the input of \textsf{Conv} (see line 5 in Algorithm~\ref{alg:reconv}). The complexity before \textsf{Conv} is summarized in Table~\ref{complexity}. Since the complexity of $\textsf{Enc}$ is larger than $\textsf{Dec}$~\cite{boemer2020mp2ml}, the cost of more than $\sigma$ \textsf{Enc} in~\cite{rathee2020cryptflow2} is more expensive than that of our $\sigma$ \textsf{Dec}. On the other hand, less than $(\sigma+d)$ \textsf{Add} are additionally required in our protocol which have negligible complexity
overhead as the \textsf{Add} is much cheaper than other PHE operations. Meanwhile, we only need half a communication round before \textsf{Conv} while~\cite{rathee2020cryptflow2} involves extra four rounds. Furthermore, the extra $2\sigma$ \textsf{Mult} in our protocol are offset for \textsf{Conv} computation to be discussed in the following.

As for complexity of \textsf{Conv}, given the encrypted \textsf{ReLU} as input,
$\mathcal{S}$ in~\cite{rathee2020cryptflow2} involves about $dc_o$ \textsf{Mult}, $dc_o$ \textsf{Add} and more than $\{(f_hf_w-1)\sigma+c_o-\left \lceil \frac{c_o}{\left \lfloor \frac{n}{h_iw_i} \right \rfloor} \right \rceil\}$ \textsf{Rot} to enable $\mathcal{C}$ to finally obtain the share of \textsf{Conv} with $\left \lceil \frac{c_o}{\left \lfloor \frac{n}{h_iw_i} \right \rfloor} \right \rceil$ \textsf{Dec}. By comparison, our joint computation eliminates the \textsf{Rot} with similar amount of \textsf{Mult} and \textsf{Add}, as well as $c_o$ \textsf{Dec} (see lines 5 and 6 in Algorithm~\ref{alg:reconv}). The complexity is also summarized in Table~\ref{complexity}. On the one
hand, if the needed $(c_o-\left \lceil \frac{c_o}{\left \lfloor \frac{n}{h_iw_i} \right \rfloor} \right \rceil)$ \textsf{Rot} in~\cite{rathee2020cryptflow2} are replaced by the same amount of \textsf{Dec}, it becomes the \textsf{Dec} complexity of our protocol. On the other hand, given the involved $2\sigma$ \textsf{Mult} before \textsf{Conv} in our protocol, the another $(f_hf_w-1)\sigma$ \textsf{Rot} for \textsf{Conv} in~\cite{rathee2020cryptflow2} is larger than $2\sigma$ as long as the kernel sizes $f_h$ and $f_w$ are larger than one. Since \textsf{Dec} and \textsf{Mult} is cheaper than \textsf{Rot}, and $f_h$ and $f_w$ are mainly three or more in many mainstream neural networks~\cite{krizhevsky2012imagenet,simonyan2014very}, our $2\sigma$ \textsf{Mult} are cheapter than $(f_hf_w-1)\sigma$ \textsf{Rot} in~\cite{rathee2020cryptflow2}.
As such, we numerically demonstrates the computation advantage of proposed $\Pi_{\textsf{ReConv}}^{\textrm{ring},p}$ over the state-of-the-art framework. Furthermore, $\mathcal{S}$ needs to send $c_o$ ciphertext to $\mathcal{C}$ in our protocol, which is about ${\left \lfloor \frac{n}{h_iw_i} \right \rfloor}$ times more than that in~\cite{rathee2020cryptflow2}. This overhead is offset by the $2c_ih_iw_i(\kappa+2\left \lceil \log{p} \right \rceil)$ bits and extra round cost before \textsf{Conv} required in~\cite{rathee2020cryptflow2}, as well as our further optimization to be elaborated in Sec.~\ref{further_opt}.

\begin{algorithm}[tb]
%\SetAlgoRefName{} % no count alg. num
\caption{Ideal functionality of \textsf{DReLU}, $\mathcal{F}_{\textsf{DReLU}}$}
\label{alg:ideal_drelu}
\SetKwInput{KwInput}{Input}
\SetKwInput{KwOutput}{Output}
\DontPrintSemicolon
\KwInput{$\langle{\bm{a}}\rangle_p^{\mathcal{C}}$ from $\mathcal{C}$, and $\langle{\bm{a}}\rangle_p^{\mathcal{S}}$ from $\mathcal{S}$ where $\bm{a}$ $\in{\mathbb{Z}_{p}^{c_i\times{h_i}\times{w_i}}}$.}
\KwOutput{$\mathcal{C}$ and $\mathcal{S}$ get $\langle{\hat{\bm{a}}}\rangle_p^{\mathcal{C}}$ and $\langle{\hat{\bm{a}}}\rangle_p^{\mathcal{S}}$, respectively, where $\hat{\bm{a}}$ $=\textsf{DReLU}(\bm{a})\in{\mathbb{Z}_{2}^{c_i\times{h_i}\times{w_i}}}$.}
\end{algorithm}
\begin{algorithm}[tb]
%\SetAlgoRefName{} % no count alg. num
\caption{Ideal functionality of \textsf{ReLU+Conv}, $\mathcal{F}_{\textsf{ReConv}}$}
\label{alg:ideal_reconv}
\SetKwInput{KwInput}{Input}
\SetKwInput{KwOutput}{Output}
\DontPrintSemicolon
\KwInput{$\langle{\bm{a}}\rangle_p^{\mathcal{C}}$ from $\mathcal{C}$, and $\langle{\bm{a}}\rangle_p^{\mathcal{S}}$, $\textrm{\textbf{K}}\in\mathbb{Z}_{p}^{c_o\times{c_i}\times{f_h}\times{f_w}}$ from $\mathcal{S}$ where $\bm{a}$ $\in{\mathbb{Z}_{p}^{c_i\times{h_i}\times{w_i}}}$.}
\KwOutput{$\mathcal{C}$ and $\mathcal{S}$ get $\langle{\bm{y}}\rangle_p^{\mathcal{C}}$ and $\langle{\bm{y}}\rangle_p^{\mathcal{S}}$, respectively, where $\bm{y}=\textsf{Conv}(\textsf{ReLU}(\bm{a}),\textrm{\textbf{K}})\in{\mathbb{Z}_{p}^{c_o\times{h_i'}\times{w_i'}}}$.}
\end{algorithm}
\noindent\textbf{3.2.5 Security Analysis.}
we now prove the semi-honest security of $\Pi_{\textsf{ReConv}}^{\textrm{ring},p}$ in the $\{$ $\mathcal{F}_{\textsf{DReLU}}$ (as shown in Algorithm~\ref{alg:ideal_drelu}) $\}$-hybrid model.

\noindent\textbf{Theorem 1.} \textit{The protocol} $\Pi_{\textsf{ReConv}}^{\textrm{ring},p}$ \textit{presented in Algorithm~\ref{alg:reconv} securely realizes} $\mathcal{F}_{\textsf{ReConv}}$ \textit{(as shown in Algorithm~\ref{alg:ideal_reconv}) in the} $\{\mathcal{F}_{\textsf{DReLU}}\}$-\textit{hybrid model, given the presence of semi-honest adversaries.}

\textit{Proof.} We construct $\textsf{Sim}_{\mathcal{C}}$ and $\textsf{Sim}_{\mathcal{S}}$ to simulate the views of corrupted client $\mathcal{C}$ and corrupted server $\mathcal{S}$ respectively.

\textbf{Corrupted client.} Simulator $\textsf{Sim}_{\mathcal{C}}$ simulates a real execution in which the client $\mathcal{C}$ is corrupted by the semi-honest adversary $\mathcal{A}$. $\textsf{Sim}_{\mathcal{C}}$ obtains $\left \langle \bm{a} \right \rangle_p^{\mathcal{C}}$ from $\mathcal{A}$, externally sends it to $\mathcal{F}_{\textsf{ReConv}}$ and waits for the output from $\mathcal{F}_{\textsf{ReConv}}$. Meanwhile, $\textsf{Sim}_{\mathcal{C}}$ waits for the $\{\textsf{ct}^{\mathcal{C}}_{\textbf{H}_{j,:}}\}_j$ from $\mathcal{A}$. $\textsf{Sim}_{\mathcal{C}}$ constructs another $\overline{\textbf{H}}$ and $\widetilde{\textbf{H}}$ with all zeros and all ones respectively, chooses the key $\textsf{pk}_{\textsf{Sim}}$ and encrypts the new $\overline{\textbf{H}}$ and $\widetilde{\textbf{H}}$ to $\{\textsf{ct}^{\textsf{Sim}}_{\overline{\textbf{H}}_{\nu,:}}\}_{\nu}$ and $\{\textsf{ct}^{\textsf{Sim}}_{\widetilde{\textbf{H}}_{\nu,:}}\}_{\nu}$ which are sent to $\mathcal{A}$. $\textsf{Sim}_{\mathcal{C}}$ randomly chooses a new $\langle{\hat{\bm{a}}}\rangle_2^{\mathcal{C}}\scriptstyle\overset{\$}{\gets}\,\mathbb{Z}_{2}^{c_i\times{h_i}\times{w_i}}$ as share, simulates $\mathcal{F}_{\textsf{DReLU}}$ sending it to $\mathcal{A}$, and waits $\{\textsf{ct}^{\textsf{Sim}}_{\acute{\textbf{H}}_{\nu,:}}\}_{\nu}$ from $\mathcal{A}$. After obtaining the output $\left \langle \bm{y} \right \rangle_p^{\mathcal{C}}$ from $\mathcal{F}_{\textsf{ReConv}}$, $\textsf{Sim}_{\mathcal{C}}$ maps it to $\left \langle \tilde{\bm{y}} \right \rangle_p^{\mathcal{C}}$ via $\phi$ (obtained from $\mathcal{A}$) and splits each column $\left \langle \tilde{\bm{y}}_{:,\beta\in[c_o]} \right \rangle_p^{\mathcal{C}}$ to the sum of $\chi$ random vectors each with size $h_i'w_i'$. Then $\textsf{Sim}_{\mathcal{C}}$ concatenates each group of those $\chi$ random vectors to form a new $\textbf{C}_{\beta}$ and encrypts it to $\textsf{ct}^{\mathcal{C}}_{\textbf{C}_{\beta}}$ via $\textsf{pk}_{\mathcal{C}}$ (obtained from $\mathcal{A}$). After that $\textsf{Sim}_{\mathcal{C}}$ sends $\{\textsf{ct}^{\mathcal{C}}_{\textbf{C}_{\beta}}\}_{\beta}$ to $\mathcal{A}$.

We argue that the output of $\textsf{Sim}_{\mathcal{C}}$ are indistinguishable from the real view of $\mathcal{C}$ by the following hybrids:

$\textsf{Hyb}_0$: $\mathcal{C}$'s view in the real protocol.

$\textsf{Hyb}_1$: Same as $\textsf{Hyb}_0$ except that the $\{\textsf{ct}^{\mathcal{S}}_{\overline{\textbf{H}}_{\nu,:}}\}_{\nu}$ and $\{\textsf{ct}^{\mathcal{S}}_{\widetilde{\textbf{H}}_{\nu,:}}\}_{\nu}$ are replaced by $\{\textsf{ct}^{\textsf{Sim}}_{\overline{\textbf{H}}_{\nu,:}}\}_{\nu}$ and $\{\textsf{ct}^{\textsf{Sim}}_{\widetilde{\textbf{H}}_{\nu,:}}\}_{\nu}$ constructed by $\textsf{Sim}_{\mathcal{C}}$. The security of PHE guarantees the view in simulation is computationally indistinguishable from the view in the real protocol.

$\textsf{Hyb}_2$: Same as $\textsf{Hyb}_1$ except that $\langle{\hat{\bm{a}}}\rangle_2^{\mathcal{C}}$ is replaced by the randomly-chosen one from $\textsf{Sim}_{\mathcal{C}}$. The security protocol $\mathcal{F}_{\textsf{DReLU}}$~\cite{rathee2020cryptflow2} guarantees the view in simulation is computationally indistinguishable from the view in the real protocol.

$\textsf{Hyb}_3$: Same as $\textsf{Hyb}_2$ except that the $\{\textsf{ct}^{\mathcal{C}}_{\textbf{C}_{\beta}}\}_{\beta}$ is replaced by the ones constructed by $\textsf{Sim}_{\mathcal{C}}$. The randomness of $\textbf{C}_{\beta}$ guarantees the view in simulation is computationally indistinguishable from the view in the real protocol and the $\left \langle \bm{y} \right \rangle_p^{\mathcal{C}}$ in real world is identical to that in
simulation. The hybrid is the view output by $\textsf{Sim}_{\mathcal{C}}$.

\textbf{Corrupted server.} Simulator $\textsf{Sim}_{\mathcal{S}}$ simulates a real execution in which the $\mathcal{S}$ is corrupted by the semi-honest adversary $\mathcal{A}$.
$\textsf{Sim}_{\mathcal{S}}$ obtains $\langle{\bm{a}}\rangle_p^{\mathcal{S}}$ and  $\textrm{\textbf{K}}$ from $\mathcal{A}$ and forwards them to the ideal functionality $\mathcal{F}_{\textsf{ReConv}}$. $\textsf{Sim}_{\mathcal{S}}$ constructs a new \textbf{H} filling with zeros, chooses the key $\textsf{pk}_{\textsf{Sim}}$, encrypts the new \textbf{H} to $\{\textsf{ct}^{\textsf{Sim}}_{\textbf{H}_{j,:}}\}_{j\in[d]}$ and sends them to $\mathcal{A}$. $\textsf{Sim}_{\mathcal{S}}$ waits for the $\{\textsf{ct}^{\mathcal{S}}_{\overline{\textbf{H}}_{\nu,:}}\}_{\nu}$ and $\{\textsf{ct}^{\mathcal{S}}_{\widetilde{\textbf{H}}_{\nu,:}}\}_{\nu}$ from $\mathcal{A}$. As for the $\mathcal{F}_{\textsf{DReLU}}$, $\mathcal{A}$ does not need to obtain output from it, and thus $\textsf{Sim}_{\mathcal{S}}$ does nothing. $\textsf{Sim}_{\mathcal{S}}$ constructs a new $\acute{\textbf{H}}$ filling with random numbers, encrypts it to $\{\textsf{ct}^{\mathcal{S}}_{\acute{\textbf{H}}_{\nu,:}}\}_{\nu}$ via $\textsf{pk}_{\mathcal{S}}$ (obtained from $\mathcal{A}$) and sends them to $\mathcal{A}$. $\textsf{Sim}_{\mathcal{S}}$ waits the $\{\textsf{ct}^{\textsf{Sim}}_{{\textbf{C}}_{\beta}}\}_{\beta}$ from $\mathcal{A}$.

We argue that the output of $\textsf{Sim}_{\mathcal{S}}$ are indistinguishable from the view of $\mathcal{S}$ by the following hybrids:

$\textsf{Hyb}_0$: $\mathcal{S}$'s view in the real protocol.

$\textsf{Hyb}_1$: Same as $\textsf{Hyb}_0$ except that $\{\textsf{ct}^{\mathcal{S}}_{\textbf{H}_{j,:}}\}_{j\in[d]}$ is replaced by the $\textsf{Sim}_{\mathcal{S}}$-constructed
$\{\textsf{ct}^{\textsf{Sim}}_{\textbf{H}_{j,:}}\}_{j\in[d]}$. The security of PHE guarantees the view in simulation is computationally indistinguishable from the view in the real protocol.

$\textsf{Hyb}_2$: Same as $\textsf{Hyb}_1$ except that the $\{\textsf{ct}^{\mathcal{S}}_{\acute{\textbf{H}}_{\nu,:}}\}_{\nu}$ is replaced by the ones constructed by $\textsf{Sim}_{\mathcal{S}}$. The randomness of $\acute{\textbf{H}}$ guarantees the view in simulation is computationally indistinguishable from the view in the real protocol. The hybrid is the view output by $\textsf{Sim}_{\mathcal{S}}$.

\begin{table}[tb]
\centering
\scriptsize
\caption{Complexity comparison for first \textsf{Conv}.}\label{complexity:iconv}
\vspace*{-0.15in}
\begin{tabular}{||c|c c c c c||}
\hline
{Framework} & \# \textsf{Rot} & \# \textsf{Enc} & \# \textsf{Mult} & \# \textsf{Dec}& \# \textsf{Add}\\
\hline
\cite{rathee2020cryptflow2}& $\geq\varpi$& $\left \lceil \frac{c_i}{\left \lfloor \frac{n}{h_iw_i} \right \rfloor} \right \rceil$ & $\approx{dc_o}$ & $\left \lceil \frac{c_o}{\left \lfloor \frac{n}{h_iw_i} \right \rfloor} \right \rceil$ &$\approx{dc_o}$\\
\hline
Ours& 0& $d$& $dc_o$ & $c_o$ & $dc_o$\\
\hline
\multicolumn{6}{l}{$\varpi=(f_hf_w-1)\sigma+c_o-\left \lceil \frac{c_o}{\left \lfloor \frac{n}{h_iw_i} \right \rfloor} \right \rceil$.}\\
\end{tabular}
\end{table}
\subsection{Further Optimizations}\label{further_opt}
\begin{algorithm}[tb]
%\SetAlgoRefName{} % no count alg. num
\caption{Compute the \textsf{Conv} with $\mathcal{C}$'s input, $\Pi_{\textsf{iConv}}^{\textrm{ring},p}$}
\label{alg:iconv}
\SetKwInput{KwInput}{Input}
\SetKwInput{KwOutput}{Output}
\DontPrintSemicolon
\KwInput{Input ${\bar{\bm{a}}}$ from $\mathcal{C}$, and $\textrm{\textbf{K}}\in\mathbb{Z}_{p}^{c_o\times{c_i}\times{f_h}\times{f_w}}$ from $\mathcal{S}$ where $\bar{\bm{a}}$ $\in{\mathbb{Z}_{p}^{c_i\times{h_i}\times{w_i}}}$.}
\KwOutput{$\mathcal{C}$ and $\mathcal{S}$ get $\langle{\bm{y}}\rangle_p^{\mathcal{C}}$ and $\langle{\bm{y}}\rangle_p^{\mathcal{S}}$, respectively, wh -ere $\bm{y}$ $=\textsf{Conv}(\bar{\bm{a}},\textrm{\textbf{K}})\in{\mathbb{Z}_{p}^{c_o\times{h_i'}\times{w_i'}}}$.}
\tcc{\underline{\textbf{offline computation}}}
$\mathcal{S}$ maps \textbf{K} to $\overline{\textbf{K}}$, gets $\{\bm{r}^{\mathcal{S}}_{\beta}{\scriptstyle\overset{\$}{\gets}}\,\mathbb{Z}_p^{n}\}_{\beta}$ and computes $\left \langle \bm{y} \right \rangle_p^{\mathcal{S}}$ by Eq.~\ref{eq:y_sp}.\;
\tcc{\underline{\textbf{$\mathcal{C}$ gets share of} $\bm{y}=\textsf{Conv}(\bar{\bm{a}},\textrm{\textbf{K}})$}}
$\mathcal{C}$ maps $\bar{\bm{a}}$ to \textbf{A} via $\iota$ and encrypts it to $\{\textsf{ct}_{\textbf{A}_{j,:}}^{\mathcal{C}}\}_j$, whi- ch are sent to $\mathcal{S}$.\;
$\mathcal{S}$ obtains $\{\textsf{ct}^{\mathcal{C}}_{\textbf{B}_{\beta}}\}_{\beta\in[c_o]}$ and then computes $\{\textsf{ct}^{\mathcal{C}}_{\textbf{C}_{\beta}}\}_{\beta}$ w -hich are sent to $\mathcal{C}$.\;
$\mathcal{C}$ decrypts $\{\textsf{ct}^{\mathcal{C}}_{\textbf{C}_{\beta}}\}_{\beta}$ to $\{\textbf{C}_{\beta}\}_{\beta}$, and gets $\left \langle \bm{y} \right \rangle_p^{\mathcal{C}}$ based on Eq.~\ref{eq:ycp}.
\end{algorithm}
\noindent\textbf{3.3.1 Deal with $\mathcal{C}$'s Input.}
Once $\mathcal{C}$ has a query to get the output of neural network, her input is firstly feed to \textsf{Conv} rather than \textsf{ReLU}. Therefore we address the \textsf{Conv} with $\mathcal{C}$'s input by $\Pi_{\textsf{iConv}}^{\textrm{ring},p}$ which is derived from the computation for \textsf{Conv} in $\Pi_{\textsf{ReConv}}^{\textrm{ring},p}$ (as shown in Algorithm~\ref{alg:iconv}). Specifically, $\mathcal{C}$ directly forms $\{\textsf{ct}_{\textbf{A}_{j,:}}^{\mathcal{C}}\}_{j\in[d]}$ by treating her private input as the output of \textsf{ReLU} namely $\bar{\bm{a}}$. Those $d$ ciphertext are sent to $\mathcal{S}$ which conducts $dc_o$ \textsf{Mult} and $dc_o$ \textsf{Add} to get $\{\textsf{ct}^{\mathcal{C}}_{\textbf{C}_{\beta}}\}_{\beta\in[c_o]}$ which are sent to $\mathcal{C}$.
$\mathcal{C}$ decrypts $\{\textsf{ct}^{\mathcal{C}}_{\textbf{C}_{\beta}}\}_{\beta}$ to $\{\textbf{C}_{\beta}\}_{\beta}$, and gets $\left \langle \bm{y} \right \rangle_p^{\mathcal{C}}$ based on Eq.~\ref{eq:ycp} (see lines 5 and 6 in Algorithm~\ref{alg:reconv}). The complexity is summarized in Table~\ref{complexity:iconv}. Similar with the complexity analysis for \textsf{Conv} in $\Pi_{\textsf{ReConv}}^{\textrm{ring},p}$, the $c_o$ \textsf{Dec} in our protocol is cheaper than that in~\cite{rathee2020cryptflow2} by
combining~\cite{rathee2020cryptflow2}'s $c_o-\left \lceil \frac{c_o}{\left \lfloor \frac{n}{h_iw_i} \right \rfloor} \right \rceil$ \textsf{Rot} with its \textsf{Dec}. On the other hand, if the another $(f_hf_w-1)\sigma$ \textsf{Rot} in~\cite{rathee2020cryptflow2} is replaced by equivalent amount of \textsf{Enc}, its \textsf{Enc} complexity becomes
\[
\{(f_hf_w-1)\sigma+\left \lceil \frac{c_i}{\left \lfloor \frac{n}{h_iw_i} \right \rfloor} \right \rceil\}\approx{d},
\]
which is the \textsf{Enc} complexity of ours. As the \textsf{Enc} is relatively cheaper than \textsf{Rot}, we thus have numerical advantage for computing the first \textsf{Conv} with $\mathcal{C}$'s input.

\noindent\textbf{3.3.2 Compute the \textsf{MaxPool} before \textsf{ReLU}.}
The \textsf{MaxPool} always appears as a form of \textsf{ReLU+MaxPool+Conv} and thus we propose to convert the \textsf{ReLU+MaxPool+Conv} to \textsf{MaxPool+ReLU+Conv} where the proposed $\Pi_{\textsf{ReConv}}^{\textrm{ring},p}$ is applied to compute \textsf{ReLU+Conv} and the \textsf{MaxPool} is calculated by $\Pi_{\textsf{tMaxPool}}^{\textrm{ring},p}$ to be described in Sec. 3.3.5. On the one hand, the correctness is guaranteed by
\[
\textsf{ReLU}(\textsf{MaxPool}(\bm{a}))=\textsf{MaxPool}(\textsf{ReLU}(\bm{a}))
\]
where $\bm{a}\in{\in{\mathbb{Z}_{p}^{c_i\times{h_i}\times{w_i}}}}$. On the other hand, we also enjoy the efficiency benefit from this conversion. Specifically, we denote the comparison process between two numbers as \textsf{Comp} and the pooling size of \textsf{MaxPool} as $s_{\textsf{mxp}}\in\mathbb{N}^+$. In the $\textsf{ReLU}(\bm{a})$, there are $c_ih_iw_i$ \textsf{Comp} calls for \textsf{DReLU}$(\bm{a})$ where the dimension of $\textsf{ReLU}(\bm{a})$ is $c_i\times{h_i}\times{w_i}$. Since $(s_{\textsf{mxp}}^2-1)$ \textsf{Comp} calls are needed to compute the subsequent \textsf{MaxPool} in each $s_{\textsf{mxp}}\times{s_{\textsf{mxp}}}$ block of $\textsf{ReLU}(\bm{a})$, the total \textsf{Comp} calls for $\textsf{MaxPool}(\textsf{ReLU}(\bm{a}))$ is
\[
c_ih_iw_i+\left \lceil \frac{h_i}{s_{\textsf{mxp}}} \right \rceil\left \lceil \frac{w_i}{s_{\textsf{mxp}}} \right \rceil(s_{\textsf{mxp}}^2-1).
\]

In contrast, while $\left \lceil \frac{h_i}{s_{\textsf{mxp}}} \right \rceil\left \lceil \frac{w_i}{s_{\textsf{mxp}}} \right \rceil(s_{\textsf{mxp}}^2-1)$ \textsf{Comp} calls are needed for $\textsf{MaxPool}(\bm{a})$, the size for each output channel becomes $\left \lceil \frac{h_i}{s_{\textsf{mxp}}} \right \rceil\times\left \lceil \frac{w_i}{s_{\textsf{mxp}}} \right \rceil$. Therefore there are $c_i\left \lceil \frac{h_i}{s_{\textsf{mxp}}} \right \rceil\left \lceil \frac{w_i}{s_{\textsf{mxp}}} \right \rceil$ \textsf{Comp} calls to compute the subsequent $\textsf{ReLU}$ and the total \textsf{Comp} calls for $\textsf{ReLU}(\textsf{MaxPool}(\bm{a}))$ becomes
\[
c_i\left \lceil \frac{h_i}{s_{\textsf{mxp}}} \right \rceil\left \lceil \frac{w_i}{s_{\textsf{mxp}}} \right \rceil+\left \lceil \frac{h_i}{s_{\textsf{mxp}}} \right \rceil\left \lceil \frac{w_i}{s_{\textsf{mxp}}} \right \rceil(s_{\textsf{mxp}}^2-1),
\]
which reduces the \textsf{Comp} calls for $\textsf{ReLU}$ computation about $s_{\textsf{mxp}}^2$ times. While this \textsf{ReLU}-\textsf{MaxPool} conversion has been applied in optimizations for GC-based and SS-based nonlinear functions~\cite{li2020falcon,liu2022securely}, or even in plaintext neural networks~\cite{relumaxpool}, they only consider the sequential  computation for the stacked functions in a neural network. We further gain the efficiency boost on top of the function-wise optimization
via our joint linear and nonlinear computation.

\noindent\textbf{3.3.3 Decouple the Mean for \textsf{MeanPool}.}
Similar with the \textsf{MaxPool}, the \textsf{MeanPool} appears as a form of \textsf{ReLU+MeanPool+Conv}. Given the summing and averaging of the \textsf{MeanPool}, we decouple these two operations such that the summing is integrated with the \textsf{ReLU} while the averaging is fused with the \textsf{Conv}. In this way, we are able to apply our $\Pi_{\textsf{ReConv}}^{\textrm{ring},p}$ to directly compute the \textsf{ReLU+MeanPool+Conv}. Specifically, we proceed the summing within the $s_{\textsf{mp}}\times{s_{\textsf{mp}}} (s_{\textsf{mp}}\in\mathbb{N}^+)$ pooling window in Eq.~\ref{eq4} such that the local addition in each $s_{\textsf{mp}}\times{s_{\textsf{mp}}}$ block is conducted from $\bm{h}_6$ to $\bm{h}_7$, which produces new $\bm{h}_6$ and $\bm{h}_7$. Meanwhile, $\mathcal{S}$ performs the same local addition for $\bm{h}_9$, and the corresponding mapping or process in Algorithm~\ref{alg:reconv} is adapted to these new matrices with shrinking size. Since each component after the above summing need to multiplied with the same number namely ${1}/{s_{\textsf{mp}}^2}$, we premultiply ${1}/{s_{\textsf{mp}}^2}$ with all components in \textbf{K} such that
\[
\textbf{K}\gets{\textbf{K}}/{s_{\textsf{mp}}^2},
\]
which achieves the averaging of \textsf{MeanPool} by averaging the subsequent kernel \textbf{K} for \textsf{Conv}. Furthermore, the $s_{\textsf{mp}}$ is usually two which results in the finite decimal 0.25 for ${1}/{s_{\textsf{mp}}^2}$.

\noindent\textbf{3.3.4 Integrate the \textsf{BN} with \textsf{Conv}.}
As the \textsf{BN} usually appears after the \textsf{Conv} and it is specified by a constant tuple $(\bm{\mu},\bm{\theta})$ in neural network inference~\cite{huang2022cheetah} where $\bm{\mu},\bm{\theta}\in\mathcal{R}^{c_o}$. Therefore, it is easily fused into the \textsf{Conv} such that
\[
\textbf{K}_{\beta,:,:,:}\gets\bm{\mu}_{\beta}{\textbf{K}_{\beta,:,:,:}}\;\textrm{and}\;\bm{b}_{\beta}\gets\bm{\theta}_{\beta}+{\bm{b}_{\beta}}.
\]

In this way, the \textsf{Conv+BN} becomes a new \textsf{Conv}, which can be combined with \textsf{ReLU} to perform our $\Pi_{\textsf{ReConv}}^{\textrm{ring},p}$ or be processed via $\Pi_{\textsf{iConv}}^{\textrm{ring},p}$.

\noindent\textbf{3.3.5 Parallel the \textsf{MaxPool} and \textsf{ArgMax}.}
Built upon the optimized \textsf{Comp}, the state-of-the-art framework sequentially conducts $(s_{\textsf{mxp}}^{2}-1)$ \textsf{Comp} calls in each $s_{\textsf{mxp}}\times{s_{\textsf{mxp}}}$ pooling window to get one component of \textsf{MaxPool}~\cite{rathee2020cryptflow2}. We further improve the efficiency via $\left \lceil \log{s_{\textsf{mxp}}^2} \right \rceil$ paralleling \textsf{Comp} calls. Specifically, the $s_{\textsf{mxp}}^{2}$ values is divided into $\left \lceil s_{\textsf{mxp}}^{2}/2\right \rceil$ pairs each have two values. Each of these pairs independently calls the \textsf{Comp} to get the maximum. The $\left \lceil s_{\textsf{mxp}}^{2}/2\right \rceil$ maximums are again divided into $\left \lceil s_{\textsf{mxp}}^{2}/2^2\right \rceil$ pairs and each pair again independently calls the \textsf{Comp} to get its maximum. The process ends with only one pair calling one \textsf{Comp} to finally get the shares of maximum among the $s_{\textsf{mxp}}^{2}$ values. Meanwhile, all of the $s_{\textsf{mxp}}\times{s_{\textsf{mxp}}}$ blocks are able to simultaneously perform $\left \lceil \log{s_{\textsf{mxp}}^2} \right \rceil$ paralleling \textsf{Comp} calls as they are mutually independent and we refer this paralleled computation as $\Pi_{\textsf{tMaxPool}}^{\textrm{ring},p}$.
On the other hand, $(s_{\textsf{amx}}-1)$ sequential \textsf{Comp} calls are involved in the \textsf{ArgMax} to get the output of a neural network~\cite{rathee2020cryptflow2} where the $s_{\textsf{amx}}$ is the number of classes in the neural network. We similarly treat the $s_{\textsf{amx}}$ values as the leaves of a binary tree such that $\left \lceil \log{s_{\textsf{amx}}} \right \rceil$ paralleling \textsf{Comp} calls are performed and we refer this paralleled process as $\Pi_{\textsf{tArgMax}}^{\textrm{ring},p}$.

\noindent\textbf{3.3.6 Performance Bonus by Channel/Layer-Wise Pruning.}
It is well-known that many mainstream neural networks have noticeable redundancy and numerous pruning techniques have been proposed to slim the networks while maintaining the model accuracy~\cite{deng2020model}. Among various pruning techniques, the channel-wise pruning contributes to reduce the number of input/output channels namely $c_i$/$c_o$ in the neural networks~\cite{he2017channel} while the layer-wise pruning deletes certain layers (i.e., removes specific functions) in the neural networks~\cite{chen2018shallowing}. Note that the above pruning does not affect the computation logic for the neural networks but with fewer functions or smaller dimensions in the functions, which serves as a performance bonus to improve the efficiency when we run the privacy-preserving protocol in the mainstream neural networks. Meanwhile, almost all the state-of-the-art frameworks rely on the intact neural networks without considering the pruned versions~\cite{liu2017oblivious,juvekar2018gazelle,mishra2020delphi,rathee2020cryptflow2,hussain2021coinn,huang2022cheetah}, and we left it as a performance bonus for practical deployment of privacy-preserving MLaaS.
%and do not apply this trick in the evaluation for a fair comparison.

\begin{table*}[!tb]
\centering
\small
\vspace*{0.05in}
\caption{Running time and communication cost of convolution layers.}\label{micro}
\begin{tabular}{||c c c c c c c c c||} \hline
Input Dimension & Kernel Dimension & \multirow{2}{*}{Stride \& Padding} & \multirow{2}{*}{Framework} & \multicolumn{2}{c}{Time (ms)} & \multicolumn{2}{c}{Speedup} & \multirow{2}{*}{Comm. (MB)}\\
$H\times{W}@c_i$ & $f_h\times{f_w}@c_o$ & & &LAN&WAN&LAN&WAN& \\
[0.5ex]
\hline\hline
\multirow{2}{*}{$14\times14@6$}&\multirow{2}{*}{$5\times5@16$} &\multirow{2}{*}{(1,\,\,0)} & \textbf{ours} & 82 & 353 & \textbf{5$\times$} & \textbf{2$\times$} & 4.39\\
& & & CrypTFlow2 & 414 & 713 &-&-& 2.2\\
\hline
\multirow{2}{*}{$2\times2@512$}&\multirow{2}{*}{$3\times3@512$} &\multirow{2}{*}{(1,\,\,1)} & \textbf{ours} &1990 &2231 & \textbf{2.5$\times$} & \textbf{2.4$\times$}& 97\\
& & & CrypTFlow2 &5173 & 5367 & -&- & 3\\
\hline
\multirow{2}{*}{$4\times4@256$}&\multirow{2}{*}{$3\times3@512$} &\multirow{2}{*}{(2,\,\,1)} & \textbf{ours} &1943 &2173 & \textbf{13$\times$}&\textbf{11$\times$} & 98\\
& & & CrypTFlow2 &25566 &25851 &- &- &5.52 \\
\hline
\multirow{2}{*}{$2\times2@512$}&\multirow{2}{*}{$1\times1@512$} &\multirow{2}{*}{(1,\,\,0)} & \textbf{ours} &1886 &2120 &\textbf{5.9$\times$} & \textbf{5.3$\times$}& 98\\
& & & CrypTFlow2 &11174 &11385 &-&-&2.6 \\
\hline
\multirow{2}{*}{$32\times32@3$}&\multirow{2}{*}{$11\times11@96$} &\multirow{2}{*}{(4,\,\,5)} & \textbf{ours} &635 &899 & \textbf{12.3$\times$}&\textbf{9.1$\times$} & 24\\
& & & CrypTFlow2 & 7866&8218 &-&-& 16\\ [1ex]
\hline
\end{tabular}
\end{table*}

\begin{table*}[!tb]
\centering
\small
\vspace*{0.15in}
\caption{Running time and communication cost on modern DL models.}\label{overall_model}
\begin{tabular}{||c | c c c c c c c c||}
\hline
Dataset& \multirow{2}{*}{DL Architecture} & \multirow{2}{*}{Framework} & \multicolumn{2}{c}{Time (ms)} & \multicolumn{2}{c}{Speedup} & \multicolumn{2}{c||}{Comm. (MB)}\\
Input Dim. ($H\times{W}@c_i$)& & &LAN&WAN&LAN&WAN&Online &Offline\\
[0.5ex]
\hline\hline
{MNIST} &\multirow{2}{*}{\texttt{LeNet}} & \textbf{ours} & 977  & 1845.5  & \textbf{ 5.3$\times$} & \textbf{ 3.3$\times$} & 52.5 &8\\
($28\times28@1$)& & CrypTFlow2 & 5243  &  6221.4 &-&-& 10.7 &-\\
\hline
 &\multirow{2}{*}{\texttt{AlexNet}} & \textbf{ours} &  35921 & 37688  & \textbf{ 2$\times$} & \textbf{ 2$\times$} & 1827 &41\\
& & CrypTFlow2 & 75196  &  76928 &-&-& 44.2 &-\\
\cline{2-9}
 &\multirow{2}{*}{\texttt{VGG-11}} & \textbf{ours} & 44366  & 47129  & \textbf{1.97$\times$} & \textbf{1.9$\times$} & 2195.3 &107.8\\
& & CrypTFlow2 & 87464  & 90359  &-&-& 168.4 &-\\
\cline{2-9}
 &\multirow{2}{*}{\texttt{VGG-13}} & \textbf{ours} & 47518  &  50749 & \textbf{1.95$\times$} & \textbf{1.89$\times$} & 2310.2 &157.3\\
& & CrypTFlow2 & 92429  &  95842 &-&-& 269.5 &-\\
\cline{2-9}
CIFAR10 &\multirow{2}{*}{\texttt{VGG-16}} & \textbf{ours} & 53499  & 57619  & \textbf{1.94$\times$} & \textbf{1.88$\times$} & 2571.9 &171\\
($32\times32@3$)& & CrypTFlow2 & 103577  &  107983 &-&-& 299.3 &-\\
\cline{2-9}
 &\multirow{2}{*}{\texttt{VGG-19}} & \textbf{ours} & 59480  & 64489  & \textbf{1.93$\times$} & \textbf{1.86$\times$} & 2833.6 &184.7\\
& & CrypTFlow2 &  114725 & 120124  &-&-& 329.1 &-\\
\cline{2-9}
 &\multirow{2}{*}{\texttt{ResNet-18}} & \textbf{ours} & 26551  &  32896 & \textbf{3.63$\times$} & \textbf{3.18$\times$} & 1306.1 &122.4\\
& & CrypTFlow2 & 96175.6  & 104308  &-&-& 267.7 &-\\
\cline{2-9}
 &\multirow{2}{*}{\texttt{ResNet-34}} & \textbf{ours} & 50360  &  63146 & \textbf{2.94$\times$} & \textbf{2.6$\times$} & 2479.7 &190.6\\
& & CrypTFlow2 &  147740.6 &  163527 &-&-& 441.8 &-\\[1ex]
\hline
\end{tabular}
\end{table*}

\begin{figure*}[!tb]
\centering
\includegraphics[trim={3.6cm 2cm 2.3cm 0cm}, clip, scale=0.58]{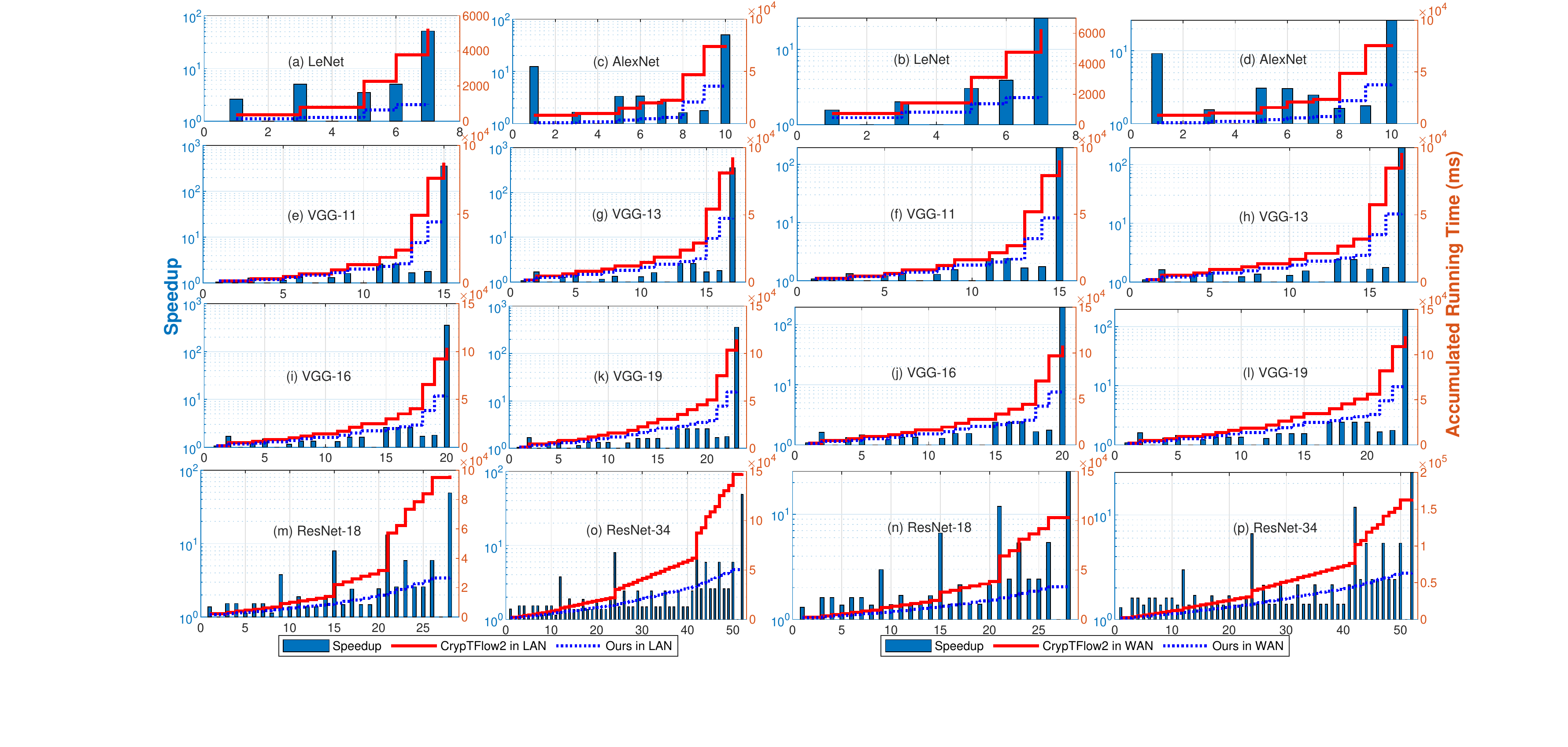}
%\vspace*{-0.15in}
\caption{Layer-wise accumulated running time and ours speedup over CrypTFlow2 on different networks: (a) and (b) LeNet; (c) and (d) AlexNet; (e) and (f) VGG-11; (g) and (h) VGG-13; (i) and (j) VGG-16;
(k) and (l) VGG-19; (m) and (n) ResNet-18; (o) and (p) ResNet-34. The x-axis denotes the layer index. The bar with values on the left y-axis indicates speedup in log scale, and the curve with values on the right y-axis indicates the accumulated
running time.
%The layers with speedup of 1 are pooling layers.
}
%\vspace*{-0.1in}
\label{time_overall}
\vspace*{-0.1in}
\end{figure*}

\begin{figure}[!tb]
\centering
\includegraphics[trim={0.3cm 2.7cm 0.6cm 0.7cm}, clip, scale=0.59]{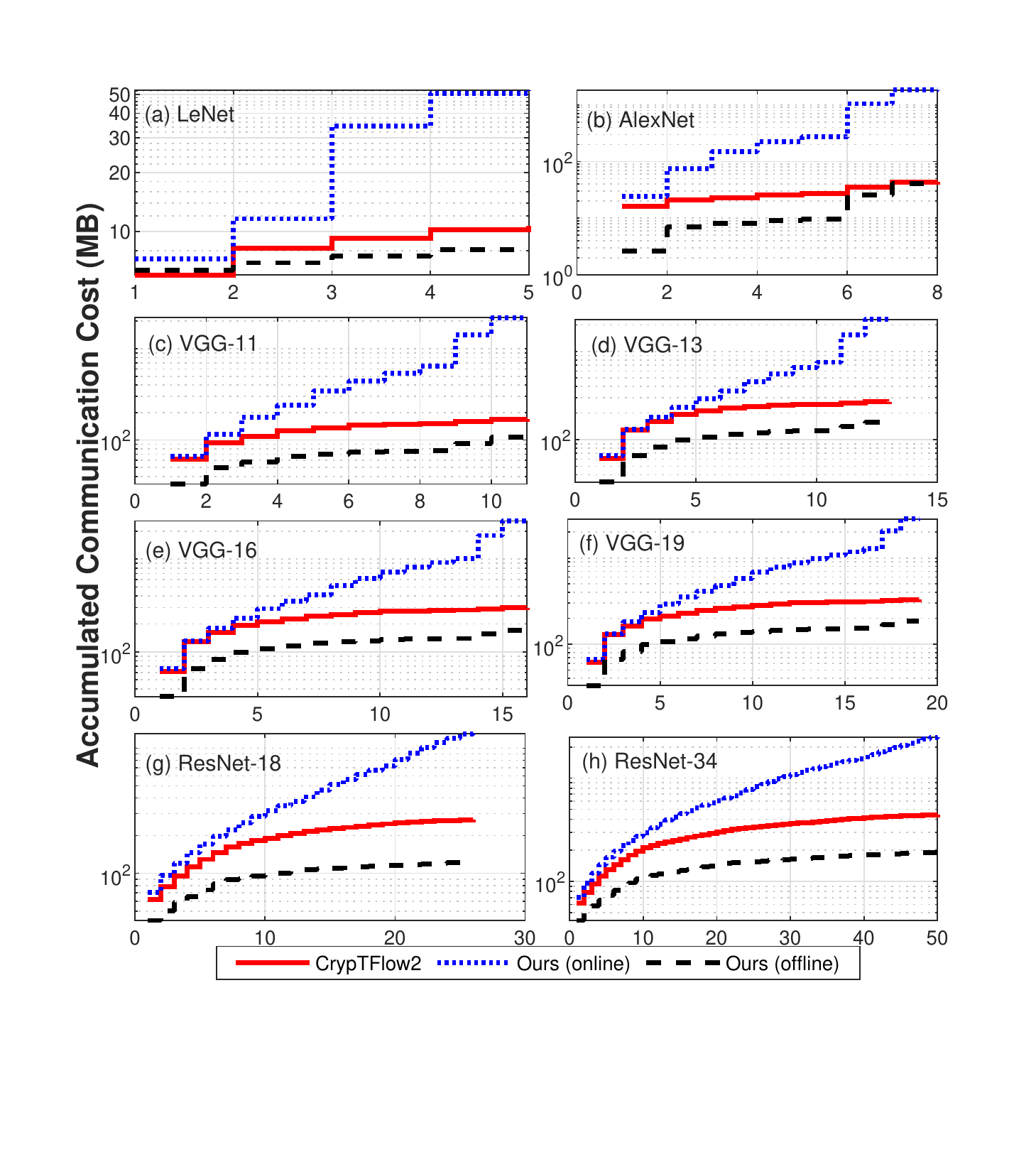}
%\vspace*{-0.15in}
\caption{Layer-wise accumulated communication cost (in log scale) on different networks: (a) LeNet; (b) AlexNet; (c) VGG-11; (d) VGG-13; (e) VGG-16;
(f) VGG-19; (g) ResNet-18; (h) ResNet-34. The x-axis denotes the layer index.}
%\vspace*{-0.1in}
\label{comm_overall}
\vspace*{-0.1in}
\end{figure}
\section{Evaluation}\label{evaluation}
In this section, we present the performance evaluation and experimental results. We first introduce the experimental setup in Sec.~\ref{sec:5:setup}, and then discuss in Sec.~\ref{microbenchmark} and \ref{online_runtime} about how efficient is our protocol to speed up the function computation and what is the prediction latency and communication cost on practical DL models by using our protocol compared with the state-of-the-art framework~\cite{rathee2020cryptflow2}\footnote{Code available at \texttt{https://github.com/mpc-msri/EzPC}.}, respectively.

\vspace*{-0.1in}\subsection{Experimental Setup}\label{sec:5:setup}\vspace*{-0.03in}
We run all experiments on two Amazon AWS
\texttt{c4.xlarge} instances possessing
the Intel(R) Xeon(R) CPU E5-2666 v3 $@$ 2.90GHz, with 7.5GB of system memory.
In the LAN setting, the client $\mathcal{C}$ and
server $\mathcal{S}$ were executed on such two instances both located in the
\texttt{us-east-1d} (Northern Virginia) availability zone. In the WAN setting, $\mathcal{C}$ and
$\mathcal{S}$ were executed on such two instances respectively located in the
\texttt{us-east-1d} (Northern Virginia)  and \texttt{us-east-2c} (Ohio) availability zone. $\mathcal{C}$ and $\mathcal{S}$ each used an 4-thread  execution. These experiential settings are similar with those used for the evaluation of the state-of-the-art frameworks~\cite{juvekar2018gazelle,mishra2020delphi}. Furthermore, we evaluate on the \textrm{MNIST}~\cite{mnist} and \textrm{CIFAR10}\cite{cifar} datasets with architectures  \texttt{LeNet}~\cite{lecun1998gradient}, \texttt{AlexNet}~\cite{krizhevsky2012imagenet}, \texttt{VGG-11}~\cite{simonyan2014very}, \texttt{VGG-13}~\cite{simonyan2014very}, \texttt{VGG-16}~\cite{simonyan2014very}, \texttt{VGG-19}~\cite{simonyan2014very}, \texttt{ResNet-18}~\cite{he2016deep}, and \texttt{ResNet-34}~\cite{he2016deep}.

%When we compare our protocol with CrypTFlow2\footnote{Code available at \texttt{https://github.com/mpc-msri/EzPC}.}, we estimate the cost of ours and CrypTFlow2 ’s protocols by summing the costs of the relevant subprotocols for all linear and non-linear transformations (including all computation and communication cost) as shown in all blocks of Figure~\ref{complexity}. We do this as a protocol decomposition with all cost counted, and this methodology is also used in other privacy-preserving frameworks.

\vspace*{-0.1in}\subsection{Microbenchmarks}\label{microbenchmark}
\vspace*{-0.03in}
We first benchmark the performance of \textsf{ReLU+Conv} in \cite{rathee2020cryptflow2} and our $\Pi_{\textsf{ReConv}}^{\textrm{ring},p}$.
In Table~\ref{micro}, we evaluate the cost of
various \textsf{ReLU+Conv} used in \texttt{LeNet}, \texttt{AlexNet}, \texttt{VGG-11}, \texttt{VGG-13}, \texttt{VGG-16}, \texttt{VGG-19}, \texttt{ResNet-} \texttt{18}, and \texttt{ResNet-34}.
The key takeaway from Table~\ref{micro} is that our online time is over 2$\times$ to 13$\times$ smaller than CrypTFlow2’s. Note that our
online communication cost is  higher than CrypTFlow2’s. However, our protocol has noticeable overall performance gain (due to the reduced computation cost as analyzed in Sec. 3.2.4 and the lower communication cost before \textsf{Conv}) to offset the total communication.

\vspace*{-0.1in}\subsection{Performance on Modern DL Models}\label{online_runtime}
\vspace*{-0.06in}
We test the performance of our protocol on various DL models used in practice. The overall evaluation is shown in Table~\ref{overall_model}. Specifically,
in the LAN setting, our protocol has a speedup of $5.3\times$, $2\times$, $1.97\times$, $1.95\times$, $1.94\times$, $1.93\times$, $3.63\times$, $2.94\times$ over CrypTFlow2 on \texttt{LeNet}, \texttt{AlexNet}, \texttt{VGG-11}, \texttt{VGG-13}, \texttt{VGG-16}, \texttt{VGG-19}, \texttt{ResNet-18}, and \texttt{ResNet-34}. The speedup is respectively $3.3\times$, $2\times$, $1.9\times$, $1.89\times$, $1.88\times$, $1.86\times$, $3.18\times$, $2.6\times$ in the WAN setting.
%As for the communication cost, the online overhead includes all transmitted data in blocks~\circled{\textit{a}}, \circled{\textit{b}}, \circled{\textit{c}}, and \circled{\textit{d}} of Figure~\ref{complexity}, while the offline overhead includes all transmitted data in offline phase as shown in Figure~\ref{complexity}.
Here we can see that our protocol has a relatively larger data load to be transmitted in the online phase compared with CrypTFlow2. As the bandwidth is a relatively low cost in today's transmission link, e.g., Amazon AWS can easily keep a bandwidth around Gigabit, the reduction of computation and communication round in our protocol brings an overall performance boost that offsets the transmission overhead.
Meanwhile, the communication overhead of our protocol in offline phase is lighter and comparable to that of CrypTFlow2 (in online phase), which has proved to be communication-reduced for the involved parties~\cite{rathee2020cryptflow2}. Furthermore, it is worth reiterating that our protocol's offline phase is \textit{totally non-interactive}, which dose not need the involved parties to synchronously exchange any calculated data. Therefore, it eliminates the interactive offline computation that is often used in the state-of-the-art frameworks such as DELPHI~\cite{mishra2020delphi} and MiniONN~\cite{liu2017oblivious}.

Next we test the runtime breakdown of each layer in our evaluated eight DL models as shown in Fig.~\ref{time_overall}, which allows us to have detailed observations.
Specifically, the runtime for each layer includes all overhead for linear and nonlinear functions. Meanwhile, the layer index also includes the pooling, e.g., mean pooling. In \texttt{LeNet}, our protocol has noticeable speedup in each convolution layer (which includes \textsf{Conv} and \textsf{ReLU}) and the speedup is larger at last layer as our protocol only needs one HE multiplication while CrypTFlow2 involves a series of HE rotations. Similar observations are found in the last layer of other networks.  In \texttt{AlexNet}, the large kernel width and height in the first layer, i.e., $f_h=f_w=11$, result in more rotations needed in CrypTFlow2 while the counterpart in our protocol is the same amount of multiplications, which is more efficient (see more details at Sec.~\ref{sys:jointcom}). At the same time, the stride of 4 in the first layer involves 16 non-stride convolutions in CrypTFlow2 while our protocol benefits from the decreased data size to be transmitted and from the smaller computation overhead for strided convolution. As such our protocol gains a larger speedup.
In \texttt{VGG-11}, \texttt{VGG-13}, \texttt{VGG-16} and \texttt{VGG-19}, the speedup is larger in layer 11, 13, 15 and 17 (except the last layer) respectively. This is because large $c_o$ (i.e., the number of output channels) makes the gap between rotation and decryption more significant. As decryption is generally cheaper than multiplication, the speedup correspondingly increases. In \texttt{ResNet-18} and \texttt{ResNet-34}, our protocol's speedup is lager in layers
9, 15, 21 (except the last layer) with strided convolution, which is in line with the speedup for first layer in \texttt{AlexNet}. Similar observations are found in the WAN setting.

Finally, we show in Figure~\ref{comm_overall} the breakdown of communication overhead in each layer in our evaluated eight DL models.
%Here the layer index doesn't include the pooling layer as no communication cost is involved in pooling.
As demonstrated in Figure~\ref{complexity}, the gap in  communication cost (i.e., the amount of data to be transmitted) between our protocol and CrypTFlow2 is proportional to the number of output channel $c_o$. As such we can see an increased difference between their communication cost as $c_o$ increases along the layers in all networks. But as we discussed earlier, despite the larger amount of data for communication, our protocol has noticeable overall performance gain (due to the reduced computation cost as analyzed in Sec.~\ref{sys:jointcom} and the lower communication cost before \textsf{Conv}) to offset the increased communication cost.

\section{Conclusions}\label{conclusion}
In this paper, we have challenged and broken the conventional compute-and-share logic for function output, and have proposed the \textit{share-in-the-middle logic}, with which we have introduced the first joint linear and nonlinear computation across functions that features by 1) \textit{the PHE triplet} for computing the intermediates of nonlinear function, with which the communication cost for multiplexing is eliminated; 2) \textit{the matrix encoding} to calculate the intermediates of linear function, with which all rotations for summing is removed; and 3) \textit{the network adaptation} to reassemble the model structure such that the recombined functions are able to take advantages of our proposed joint computation block as much as possible.
The boosted efficiency of our protocol has been verified by the numerical complexity analysis and the experiments have also demonstrated up to 13$\times$ speedup for various functions used in the state-of-the-art models and up to $5\times$ speedup over mainstream neural networks compared with the state-of-the-art privacy-preserving frameworks.  Furthermore, as the first one for share-in-the-middle computation, we have considered the joint computation between two adjacent functions while combing more functions forms an interesting work to explore more efficiency boost.

\bibliographystyle{IEEEtran}
\bibliography{ccs-sample}

%\vspace*{-0.5in}
\begin{IEEEbiography}[{\includegraphics[width=1in,height=1.25in,clip,keepaspectratio]{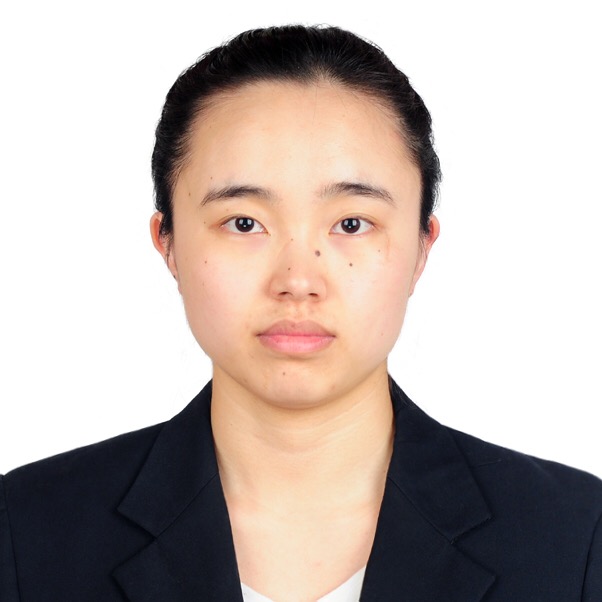}}]{Qiao Zhang}
is an assistant professor at College of Computer Science, Chongqing University, China. Before that, She obtained her Ph.D. degree at Department of Electrical and Computer Engineering in Old Dominion University (21'). Her current research is about the privacy-preserved machine learning and she has published papers such as in NDSS, AsiaCCS, IJCAI, IEEE IoT Journal, IEEE TNSE.
\end{IEEEbiography}
%\vspace*{-0.6in}
\begin{IEEEbiography}[{\includegraphics[width=1in,height=1.25in,clip,keepaspectratio]{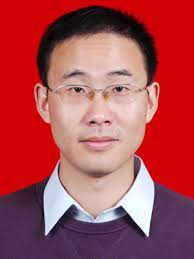}}]{Tao Xiang}
received the B.Eng., M.S. and Ph.D. degrees in computer science from Chongqing University, China, in 2003, 2005 and 2008, respectively. He is currently a Professor of the College of Computer Science at Chongqing University.
Dr. Xiang’s research interests include multimedia
security, cloud security, data privacy and cryptography. He has published over 90 papers on
international journals and conferences. He also
served as a referee for numerous international journals and conferences.
\end{IEEEbiography}
%\vspace*{-0.6in}
\begin{IEEEbiography}[{\includegraphics[width=1in,height=1.25in,clip,keepaspectratio]{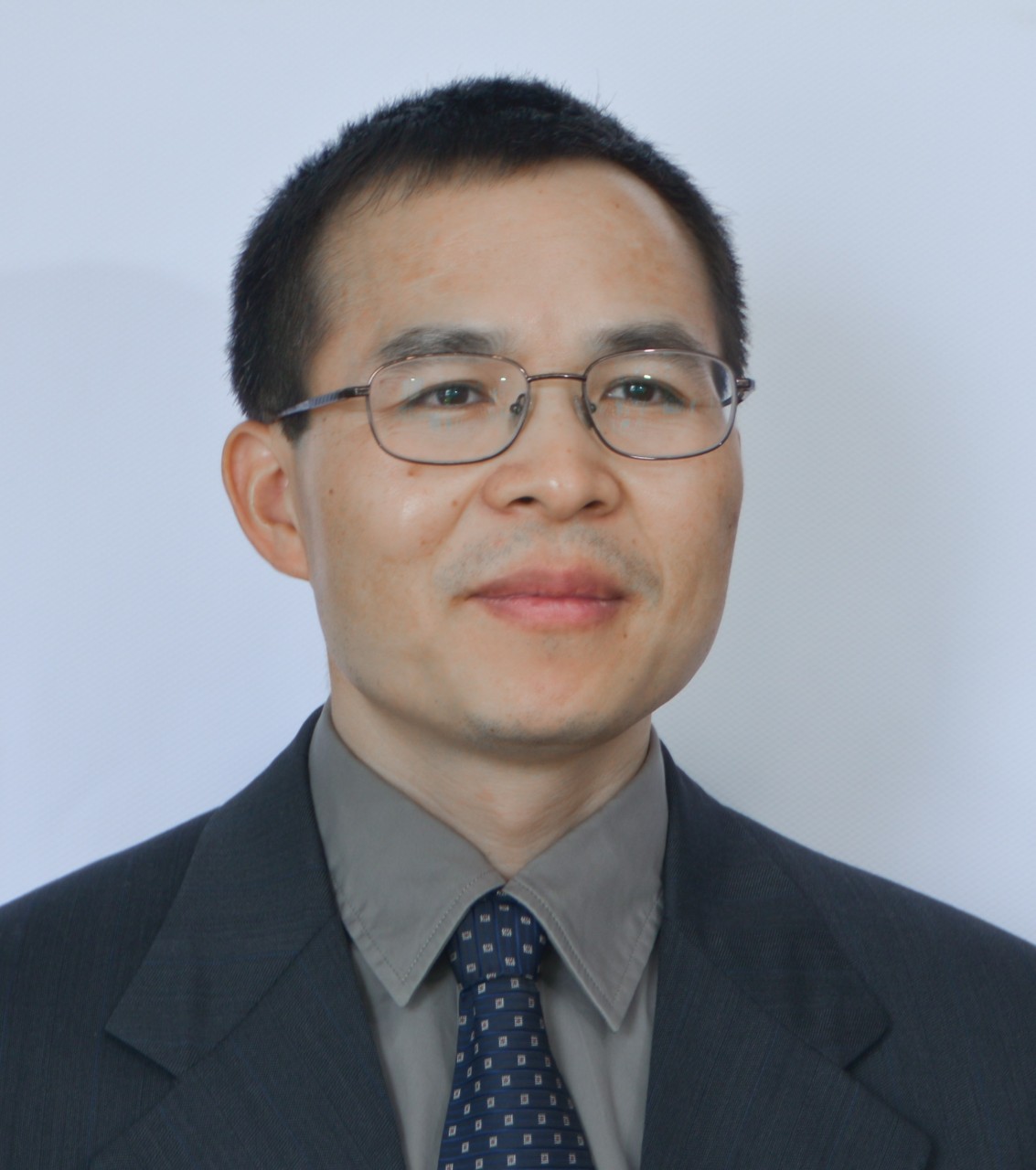}}]{Chunsheng Xin}
is a Professor in the Department of Electrical and Computer Engineering, Old Dominion University. He received his Ph.D. in Computer Science and Engineering from the State University of New York at Buffalo in 2002. His interests include cybersecurity, machine learning, wireless communications and networking, cyber-physical systems, and Internet of Things. His research has been supported by almost 20 NSF and other federal grants, and results in more than 100 papers in leading journals and conferences, including three Best Paper Awards, as well as books, book chapters, and patent. He has served as Co-Editor-in-Chief/Associate Editors of multiple international journals, and symposium/track chairs of multiple international conferences including IEEE Globecom and ICCCN. He is a senior member of IEEE.
\end{IEEEbiography}
%\vspace*{-2in}
\begin{IEEEbiography}[{\includegraphics[width=1in,height=1.25in,clip,keepaspectratio]{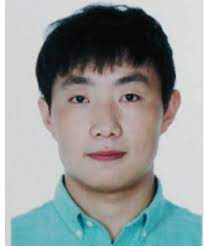}}]{Biwen Chen}
received his Ph.D. degree from School of Computer, Wuhan University in 2020. He is currently an assistant professor of the School of Computer, Chongqing University, China. His main research interests include cryptography, information security and blockchain.
\end{IEEEbiography}
%\vspace*{-2in}
\begin{IEEEbiography}[{\includegraphics[width=1in,height=1.25in,clip,keepaspectratio]{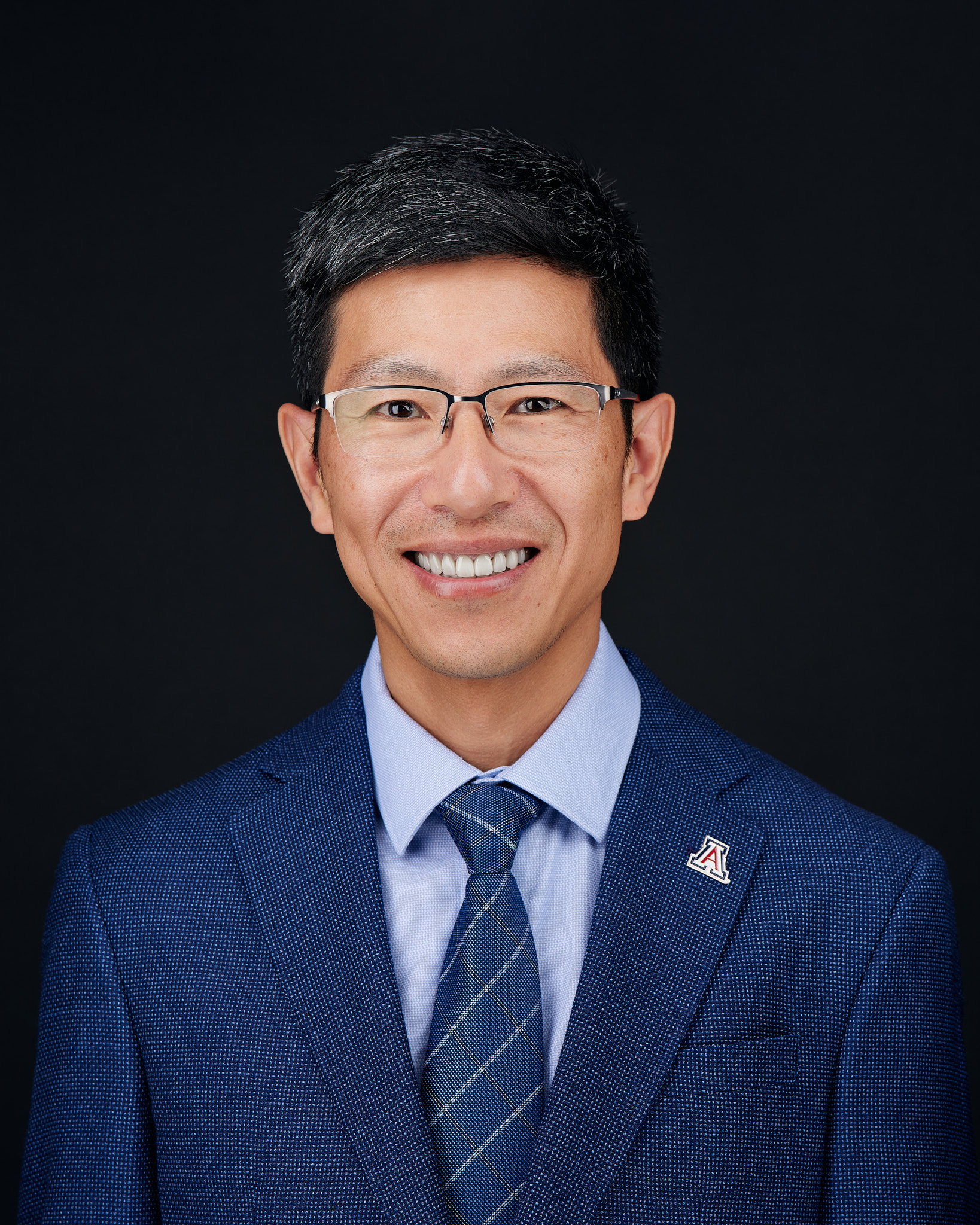}}]{Hongyi Wu}
is a Professor and Department Head of Electrical and Computer Engineering at the University of Arizona. Between 2016-2022, He was a Batten Chair Professor and Director of the School of Cybersecurity at Old Dominion University (ODU). He was also a Professor in the Department of Electrical and Computer Engineering and held a joint appointment in the Department of Computer Science. Before joining ODU, He was an Alfred and Helen Lamson Professor at the Center for Advanced Computer Studies (CACS), the University of Louisiana at Lafayette. He received a B.S. degree in scientific instruments from Zhejiang University, Hangzhou, China, in 1996, and an M.S. degree in electrical engineering and a Ph.D. degree in computer science from the State University of New York (SUNY) at Buffalo in 2000 and 2002, respectively. His current research focuses on security and privacy in intelligent computing and communication systems. He chaired several conferences, including IEEE Infocom 2020. He served on the editorial board of several journals, such as IEEE Transactions on Computers, IEEE Transactions on Mobile Computing, IEEE Transactions on Parallel and Distributed Systems, and IEEE Internet of Things Journal. He received an NSF CAREER Award in 2004, UL Lafayette Distinguished Professor Award in 2011, and IEEE Percom Mark Weiser Best Paper Award in 2018. He is a Fellow of IEEE.
\end{IEEEbiography}

\end{document}